\documentclass[manuscript]{aastex63}

\newcommand\pcc{{\rm\thinspace cm^{-3}}}
\def\ps{{\rm\thinspace s^{-1}}}
\providecommand{\e}[1]{\ensuremath{\times 10^{#1}}}
\submitjournal{\apj}
\shorttitle{Cosmic-Ray ionization rate}
\shortauthors{Shaw \& Ferland}
\begin{document}
\correspondingauthor{Gargi Shaw}
\email{gargishaw@gmail.com, gary@g.uky.edu}

\author{Gargi Shaw}

\affiliation{Department of Astronomy and Astrophysics, Tata Institute of Fundamental Research\\
Homi Bhabha Road, Navy Nagar, Colaba, Mumbai 400005, India}

\author{G. J. Ferland}

\affiliation{Department of Physics and Astronomy, University of Kentucky\\
Lexington, KY 40506, USA}

\title{Role of Polycyclic Aromatic Hydrocarbons on the Cosmic-Ray ionization rate in the Galaxy}

\begin{abstract}
The cosmic-ray ionization rate ($\zeta$, s$^{-1}$) plays an important role in the interstellar medium. 
It controls ion-molecular chemistry and provides a source of heating.
Here we perform a grid of calculations using the spectral synthesis code CLOUDY 
along nine sightlines towards, HD 169454, HD 110432, HD 204827, $\lambda$ Cep, X Per, 
HD 73882, HD 154368, Cyg OB2 5, Cyg OB2 12. 
The value of $\zeta$ is determined by matching the observed column densities of H$_3^+$  and H$_2$. 
The presence of polycyclic aromatic hydrocarbons (PAHs) affects the free electron density,
which changes the  H$_3^+$ density and
the derived ionization rate. PAHs are ubiquitous in the Galaxy, but there are also regions where 
PAHs do not exist.
Hence, we consider clouds with a range of PAH abundances and show their effects
on the  H$_3^+$ abundance. 
We predict an average cosmic-ray ionization rate for H$_2$ 
($\zeta$(H$_2$))= (7.88 $\pm$ 2.89) $\times$ 10$^{-16}$ s$^{-1}$ for models with 
average Galactic PAHs abundances, (PAH/H =10$^{-6.52}$), 
except Cyg OB2 5 and Cyg OB2 12. The value of $\zeta$ is nearly 1 dex smaller for 
sightlines toward Cyg OB2 12. 
We estimate the average value of $\zeta$(H$_2$)= (95.69 $\pm$ 46.56) $\times$ 10$^{-16}$ s$^{-1}$ 
for models without PAHs. 
\end{abstract}




\section{Introduction}\label{sec:intro}
Cosmic rays (CR) are high-energy particles mostly made of protons whose energy ranges 
from few MeV to few GeV. 
They penetrate deep into interstellar clouds and produce a primary ionization 
and a cascade of secondary ionizations \citep{1968ApJ...152..971S}. 
The free electrons they produce heat the gas, and deep in the cloud they control the ion-molecular 
chemistry network in the absence of other ionizing sources.  
Hence it is very important to determine the cosmic-ray ionization rate ($\zeta$) for
such environments. Several groups estimated the value of $\zeta$ based on the 
abundances of various molecules 
such as HD \citep{1974ApJ...191...89O}, H$_3^+$ \citep{{2012ApJ...745...91I},{2017ApJ...845..163N}}. 
Interestingly, the value differs depending on the molecule involved. 
In the diffuse clouds, $\zeta$ is determined from the observed column densities 
of H$_3^+$ since it undergoes simple chemical reactions \citep{2003Natur.422..500M}. 
However, H$_3^+$ undergoes a complex chain of chemical reactions in dense 
clouds \citep{2006PNAS..10312269D} and the value of $\zeta$ is lower than that of diffuse clouds. 

The current estimate of the average value of $\zeta$ (3.5$^{+5.3}_{-3.0}$ $\times$ 10$^{-16}$ s$^{-1}$) 
in the galaxy is 
provided by \citet{2003Natur.422..500M}, \citet{2007ApJ...671.1736I}, 
\citet{2012ApJ...745...91I} based on H$_3^+$ observations. Similar value 
of (5.3$\pm$1.1) $\times$ 10$^{-16}$ s$^{-1}$ has been reported by \citet{2017ApJ...845..163N}.  
Earlier to the H$_3^+$ revolution, the value of $\zeta$ had been 
derived using other molecules, such as HD and OH \citep{{1974ApJ...191...89O},{1977ApJS...34..405B},{1978MNRAS.185..643H},{1996ApJ...463..181F}}. 
However, H$_3^+$ gives a higher value than other methods, 
and the value of $\zeta$ varies by nearly an order of magnitude across the Galaxy \citet{2012ApJ...745...91I}. 
\citet{2016A&A...585A.105L} and \citet{2019ApJ...883...54O} 
have reported even higher values in the central molecular zone of the Galaxy.
In addition, \citet{2016MNRAS.459.3234S} and \citet{2018MNRAS.481.2083R} have  
reported higher value of $\zeta$ in 
high-redshift H$_2$ bearing Damped Lyman alpha absorbers.
      
The H$_3^+$ method relies on measurements of the column density of 
that molecular ion and needs the electron density in the H$_3^+$ region 
to derive $\zeta$.  
It is not possible to measure the electron density directly, so indirect assumptions, 
described in the next section, have been made.  
\citet{2006PNAS..10312269D} and \citet{2017ApJ...845..163N} have 
examined the chemical reactions which affect the H$_3^+$.
Here we examine the effects of polycyclic aromatic hydrocarbons (PAHs) on the electron density 
and determine its impact on determining the value of $\zeta$.

This paper is organized in the following manner: in Section 2,
we describe the H$_3^+$ chemistry and the role of electron density. Our grid of 
numerical simulations of nine sightlines and the results are presented in Sections 3 and 4, respectively.
Discussions and Summary are given in section 5 and 6, respectively.

\section {H$_3^+$ chemistry and electron density} \label{sec:Chemistry}

\subsection{Observations and analytical analysis}
The molecular ion H$_3^+$ is a very important species which can be used to estimate the cosmic-ray ionization rate \citep{{1998Sci...279.1910M},{2003Natur.422..500M}}. 
In the diffuse interstellar clouds, H$_3^+$ is mainly formed by the cosmic ray ionization of H$_2$,
\begin{equation}
H_2 + CR = H_2^+ + e^-
\end{equation}
followed by proton abstraction reaction producing 
\begin{equation}
H_2^+ + H_2 = H_3^+ + H.
\end{equation} 
Whereas, H$_3^+$ is destroyed mainly by dissociative recombination processes,
\begin{equation}
H_3^+ + e^- = H + H + H 
{\rm\ or\ }
H_2 + H \ .
\end{equation}
Process 2 is very slow, so in steady state we can write,
\begin{equation}
\zeta (H_2) n(H_2) = k_e n(e^-) n(H_3^+).
\end{equation}
Here, $\zeta$(H$_2$) and k$_e$ represent the cosmic-ray ionization rate of H$_2$ 
and the electron-recombination co-efficient of H$_3^+$, respectively. 
The terms $n$(X) represent the number density ($\pcc$) of species X.
For a uniform cloud with length L (cm),  
the densities can be converted into column densities $N$ (cm$^{-2}$) 
and equation 4 takes the  form
\begin{equation}
\zeta (H_2) = k_e n(e^-) N(H_3^+)/ N(H_2) \ [\ps]. \ 
\end{equation}
Defining the molecular fraction  as
\begin{equation}
f = 2N(H_2) / N(H),
\end{equation}
one gets
\begin{equation}
\zeta(H_2)=2 k_e n(_e) N (H_3^+)/ f N(H) [\ps].
\end{equation}

\citet{2012ApJ...745...91I} have observed H$_3^+$ in absorption along 
50 diffuse sightlines and derived the column densities needed to
evaluate equation 7. 
The observed column densities of H$_3^+$ are 
distributed over a range 5$\times$10$^{13}$ to 3.7$\times$10$^{14}$ cm$^{-2}$. 
An electron density is required to derive the cosmic-ray ionization rate.

\citet{2007ApJ...671.1736I} and \citet{2012ApJ...745...91I} derived $\zeta (H_2)$ by
assuming $f=0.67$. 
In reality, a large range in $f$ is possible so there
is significant uncertainty 
in estimating the values of $f$ and $n$(H).  
\citet{1986ApJS...62..109V} showed that in diffuse clouds, 
nearly all carbon is singly ionized. 
\citet{2007ApJ...671.1736I} further assumed that 
in diffuse clouds, nearly all electrons are produced via the ionization of C to C$^+$
and C/H should approximate the electron fraction, e$^-$/H, that is,
$n_e = n(C^+)$.
They assumed a single average value, 
$n_e / n(H) = 1.4 \e{-4}$,
for all of their calculations and derived a high value for $\zeta$(H$_2$) 
compared to earlier  estimates. 

\subsection{Numerical methods}
Diffuse clouds show structures like photodissociation regions (PDR) \citep{1985ApJ...291..722T}.  
Hence, we first investigate the above-mentioned assumptions regarding 
electron density for a standard PDR model. 
Then we perform detailed numerical simulations of nine sightlines, 
which are discussed later, to estimate an average cosmic-ray ionization rate.   

All the models presented here are calculated using spectroscopic simulation code, CLOUDY 
\citep{{2013RMxAA..49..137F},{2017RMxAA..53..385F},{2005ApJ...624..794S},{2020RNAAS...4...78S}}. 
For this work, we have updated the following rate coefficients 
for the reactions involving H$_3^+$,
\begin{equation}
H_3^+ + CO =  H_2 + HCO^+
\end{equation}
\begin{equation}
H_3^+ + CO =  H_2 + HOC^+
\end{equation}
\begin{equation}
H_3^+ + O = H_2 + OH^+
\end{equation}
\begin{equation}
H_3^+ + N_2 = H_2 + HN_2^+ .
\end{equation}
These rates are from \citet{2010JPCA..114..278K}, and  \citet{1982IJMSI..41..185R}, respectively. The reaction rates for equation 2 and 3 are taken from \citet{1974JChPh..60.2840T} and \citet{2004PhRvA..70e2716M}, respectively. \citet{2004PhRvA..70e2716M} provides the total dissociative recombination rate for the products (eqn.3).  
UMIST RATE12 (http://udfa.ajmarkwick.net) provides a branching ratio of 0.53 : 1. The branching ratios of any product reaction rate is important. Though a branching ratio 0.53 : 1 works for the models discussed here, it introduces instability in chemical 
networks for primordial IGM models. In CLOUDY, we use a branching ratio 0.25 : 1 . 
For other atomic, ionic, and molecular processes, CLOUDY utilises five distinct databases \citep{{2012MNRAS.425L..28P}, {2015ApJ...807..118L},{2005ApJ...624..794S}}. 

We use the size-resolved PAHs, distributed in 10 size bins, and the size distribution of the PAHs is taken from \citet{2008ApJ...686.1125A} with minimum and maximum radii of 0.00043 $\mu$m (30 C atoms) 
and 0.0011 $\mu$m (500 C atoms), respectively.  Non-equilibrium heating is important for these 
small grains, and we include this effect in our calculations. Besides this, we also consider
the photoelectric effect as well as charge exchange of PAHs with atoms/ions and electrons to determine the charge of the PAHs.  Inner-shell photoionization and the Auger effect of grains and PAHs are also treated following \citet{2006ApJ...645.1188W}. 
We include the opacity of both neutral and charged PAHs in our calculation with opacities 
according to \citet{2001ApJ...550L.213L}, who adopted a thermal approximation. Grains have a net charge, and 
so affect the density of free electrons.
Our treatment of this physics is described in \citet{2008ApJ...686.1125A}. In all the models discussed here 
(if not specified), the temperature is determined from heating 
and cooling balance involving various terms \citep{2017RMxAA..53..385F}.

\subsection{Ionization structure, electron density and Carbon to Hydrogen ratio} \label{subsec:ionization}

We test the assumptions about the electron density by running the 
standard Leiden PDR models (F1, F2, F3, F4) \citep{2007A&A...467..187R} and 
finding the predicted H$_3^+$. 
We find column densities for the F1, F2, F3, and F4 models to be 
9.8$\times$10$^{14}$, 3.29$\times$10$^{14}$, 1.4$\times$10$^{13}$, 
1.09$\times$10$^{13}$ cm$^{-2}$  respectively. 
All these models extend up to Av=10. 
However, the samples of diffuse cloud considered by \citet{2012ApJ...745...91I} extend less than Av=10. Hence as a next step, we run the F1 and F2 models 
stopping at N(H$_3^+$) = 10$^{13.8}$ cm$^{-2}$ and check 
the corresponding Av. We find Av= 3.16 and 6.9 for F1 and F2 models, respectively. Hence we consider F1 as a standard model for further investigation as many of the clouds of the \citet{2012ApJ...745...91I} sample have Av in
these ranges, and examine the basis for estimates of the electron fraction to be equal to C/H.

The model F1 is a constant temperature PDR model at 50 K with hydrogen density 10$^3$ cm$^{-3}$ and impinging radiation field 
with a strength of 17.0 G0, where G0 is the widely used standard FUV radiation field measured 
in units of 1.6$\times$10$^{~3}$ ergs cm $^{-2}$s$^{-1}$ \citep{1985ApJ...291..722T}. 
Details of this model is discussed in \citet{2007A&A...467..187R}. 
The only difference between  
\citet{2007A&A...467..187R} and this work is that here we consider 
ISM gas phase abundances 
\citep{{1986ARA&A..24..499C},{1996ARA&A..34..279S},
{1998ApJ...493..222M},{2007ApJ...655..285S}}, 
$\zeta$(H)= 2$\times$ 10$^{-16}$ s$^{-1}$,  
and extend the model till N(H$_3^+$) = 10$^{13.8}$ cm$^{-2}$. 
Our work does not use the simplified expression for electron density as 
suggested by \citet{2012ApJ...745...91I}. Instead, we use a detailed chemical network 
with appropriate microphysics \citep{{2008ApJ...686.1125A},{2013RMxAA..49..137F},{2020MNRAS.493.5153S}}. 

\begin{figure}[]
\includegraphics[scale=0.7]{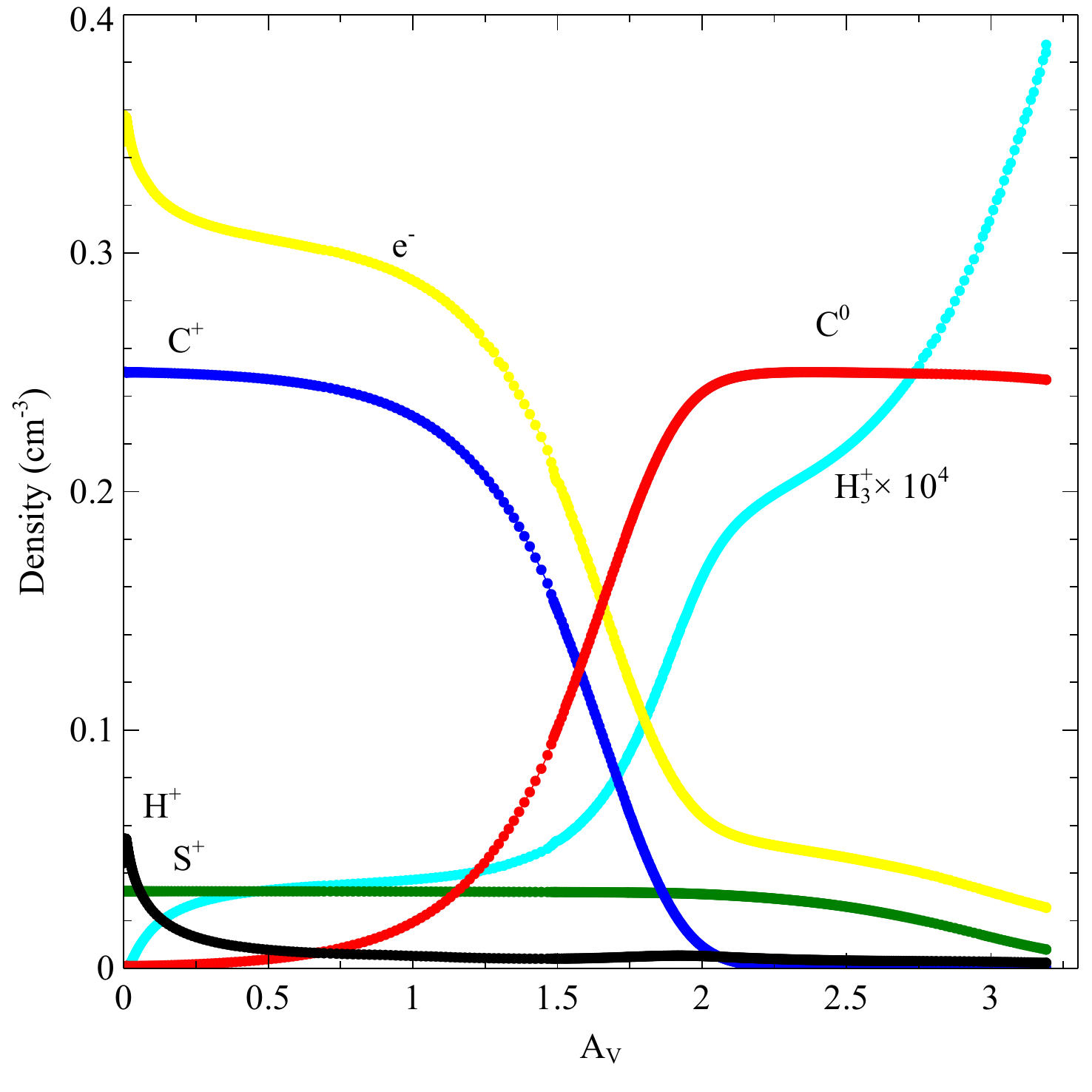}
\caption{ Ionization structure for model F1. Density of C$^0$, C$^+$, H$^+$, S$^+$, H$_3^+$, and e$^-$ are plotted as a function of Av.\label{fig:fig1}}
\end{figure}

Fig. {\ref{fig:fig1}} shows the density of C$^0$, C$^+$, H$^+$, S$^+$, H$_3^+$, and e$^-$ as a function of Av. 
It is clear from the plot that 
{\it a}) all the carbon is not in C$^+$ throughout the entire cloud. 
Carbon is mostly in the form of C$^{+}$ at shallower Av 
and decreases as Av increases. 
{\it b}) all the electrons are not contributed by C$^+$. 
Ionization of H$^+$, S$^+$, and other metals also contribute 
to the total electron density. 
Hence, 
e$^-$/H is not equal to C/H. For this model e$^-$/H $>$ C/H. 

\begin{figure}[]
\includegraphics[scale=0.7]{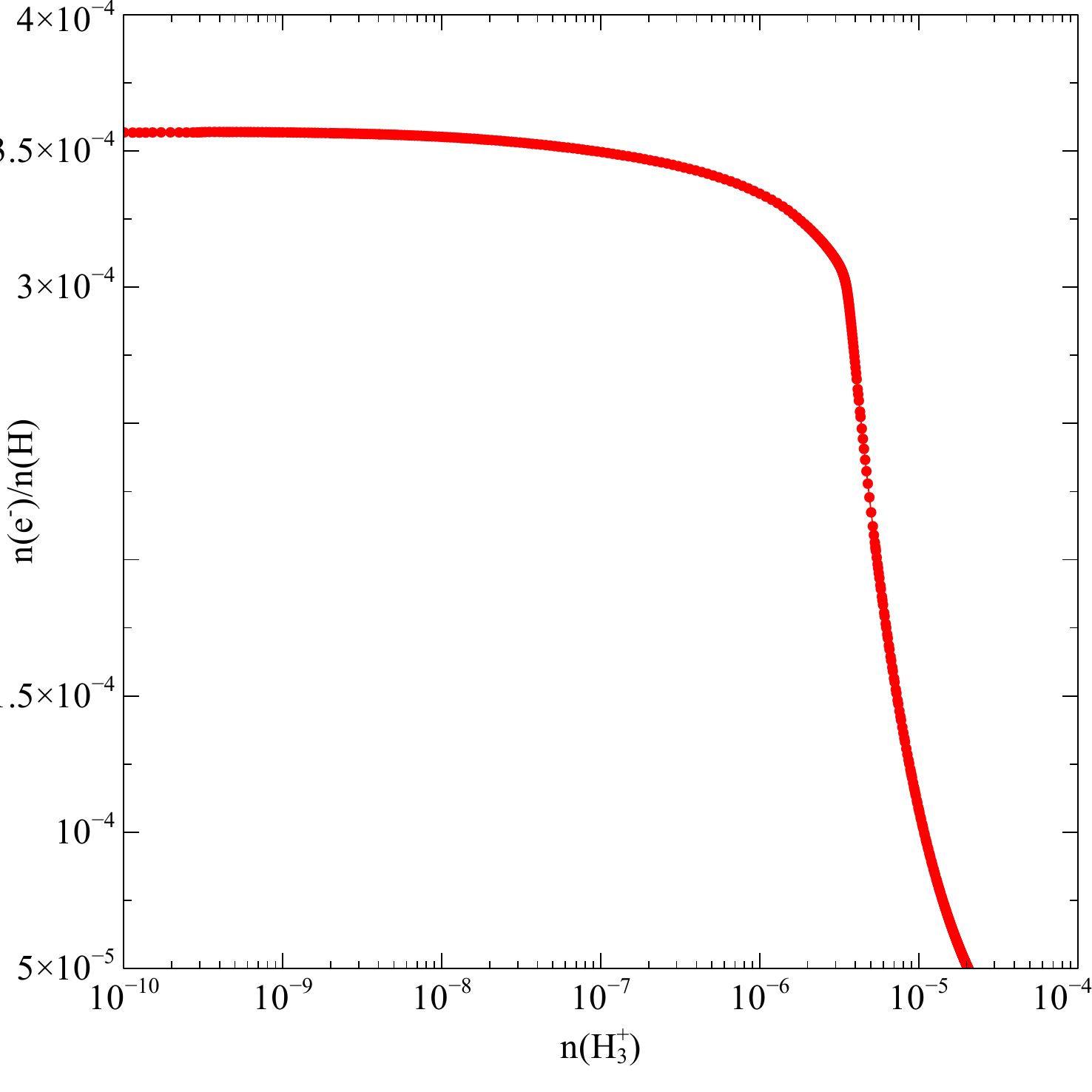}
\caption{ $n$(e$^-$)/$n$(H) is plotted as a function of $n$(H$_3^+$) for model F1. \label{fig:fig2}}
\end{figure}

Fig. {\ref{fig:fig2}} shows the electron fraction as a function of 
$n$(H$_3^+$) for this model. 
Initially $n$(e$^-$)/$n$(H) is 
higher than the assumed C/H abundance ratio of 2.5$\times$ 10$^{-4}$.
This is because ions of H, S, Mg provide electrons besides C. 
On top of this, 
the electron fraction decreases nearly by a factor 10 at higher Av where H$_3^+$ forms. 
This would have a significant impact on the estimate of $\zeta$ (see equation 7).

\subsection{Effect of physical conditions on electron density} \label{subsec:effect}
From PDR models it is known that other elements besides C also contribute to the total electron density 
depending on the density, impinging 
radiation fields and metallicity.
Additionally, grains can add or remove electrons from the gas. 
PAHs also change the electron density due to its high electron affinity \citep{2013A&A...549A.103C}. 
\citet{2009ApJ...700.1299J} has also shown that metals are depleted in 
different amounts across the Galaxy. 
On top of this, there is an additional 
positive feedback contribution of electrons from the cosmic-ray ionization of H, $\zeta$ (H), since the number 
of secondary electrons that are produced
by a primary ionization depends on the electron fraction. 
It is to be noted here that generally $\zeta$(H) = 0.5 $\times$ $\zeta$(H$_2$). 
In light of these facts, here we show the effects of various physical conditions
on electron density.
 
\begin{figure}[]
\includegraphics[scale=0.7]{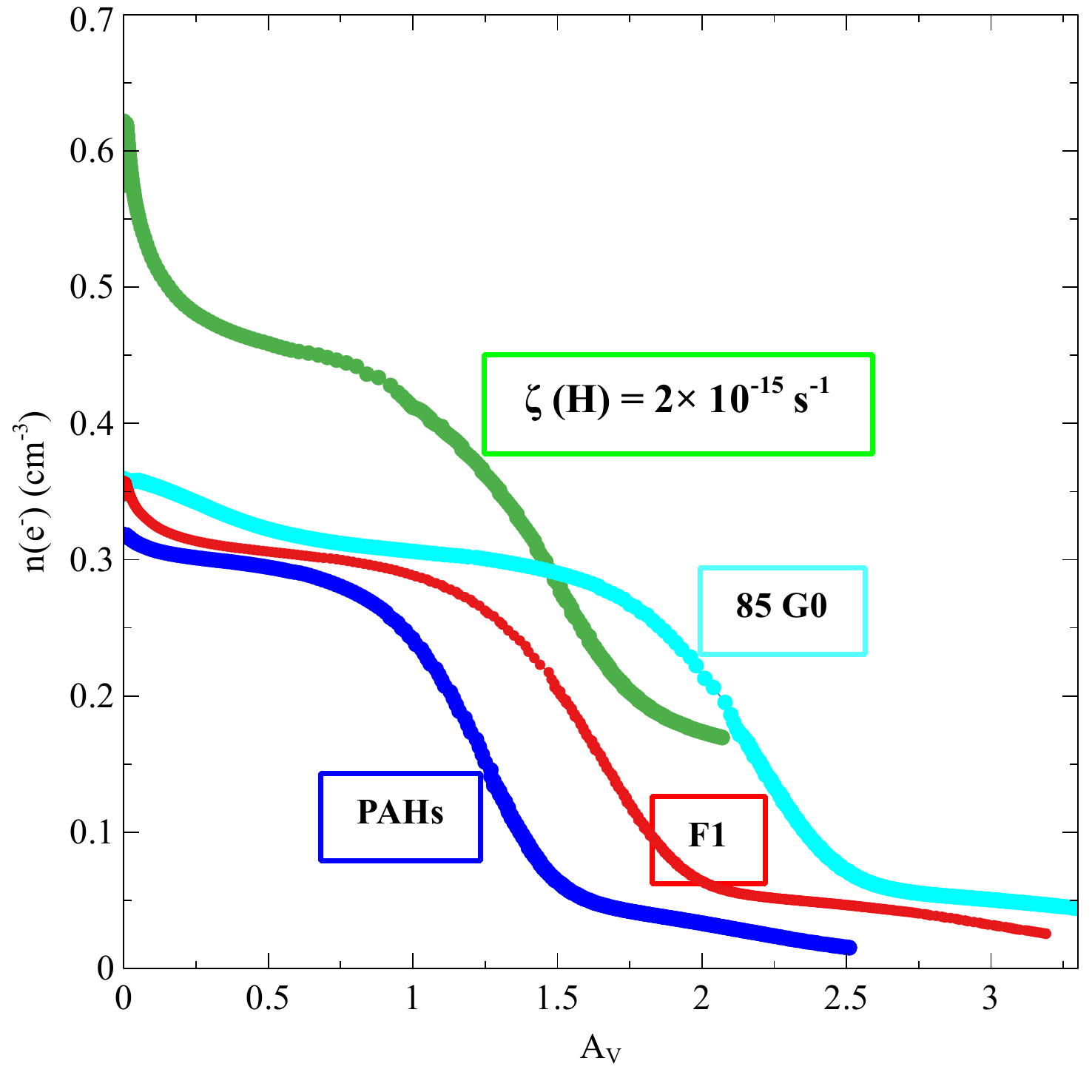}
\caption{The red, cyan, green, blue solid lines show $n$(e$^-$) as a function of Av for default F1, and with increased radiation field, increased $\zeta$(H), 
and with PAHs, respectively.
All clouds have a thickness that reproduces the same column density
in H$_3^+$ (10$^{13.8}$ cm$^{-2}$). \label{fig:fig3}}
\end{figure}

As a simple test, we ran the Leiden F1 model varying single parameters 
while keeping all the other parameters the same.
Unlike the canonical Leiden PDR model, we include PAHs.
We varied the radiation field and the cosmic-ray ionization rate
 $\zeta$. 
 Fig. {\ref{fig:fig3}} 
shows the electron densities as a function of Av with a radiation field 
with strength 85 G0 (cyan),
$\zeta$(H)= 2$\times$ 10$^{-15}$ s$^{-1}$ (green), and PAHs (blue). 
The default case is shown with the red solid line.  
For these cases, the shape of the 
$n$(e$^-$)/$n$(H) vs. $n$(H$_3^+$) plot looks similar but the 
value of $n$(e$^-$)/$n$(H) is higher. 
Earlier  \citet{2008ApJ...675..405S} 
showed that the grain physics plays an important role in determining the value of $\zeta$. 
Small grains neutralize ions and remove free electrons. 
They also extinguish the FUV radiation field, 
and hence influence the deduced value of $\zeta$. 
\citet{2012ApJ...745...91I} assumed a single average value 
(1.4 $\times$10$^{-4}$) for e$^-$/H. For model F1 the value of total hydrogen density is 10$^3$ cm$^{-3}$. 
Hence, their derived electron density 
would be 0.14 cm$^{-3}$ throughout the entire cloud. 
Considering all these facts, we revisit the value of $\zeta$ by modelling various
sight-lines with observed column densities of H$_3^+$ and H$_2$, using 
detailed numerical simulation.

\section{Grids of models } \label{sec:calculations}
Here we present numerical simulations of nine sightlines from \citet{2012ApJ...745...91I}, namely, HD 169454, HD 110432, HD 204827, $\lambda$ Cep, X Per, HD 7388, HD 154368, Cyg OB2 5, and Cyg OB2 12. 
To minimize uncertainties in modelling, we choose 
only those sightlines for which the density 
is derived from observed C$_2$ levels \citep{{2007ApJS..168...58S},{2002ApJ...577..221R}}. 
All these sightlines have reported H$_2$ observations and E(B-V). 
We assume that the classical grains have a \citet{1977ApJ...217..425M} size distribution.
Hence, we consider R=3.2 and calculate Av as 3.2$\times$ E(B-V). 
Very little is known about the presence or absence of PAHs along these sightlines.

Table {\ref{tab:1}} lists all the observed information for the nine  
sightlines modelled here. 
For each sightline, we run a grid of models with varying radiation field and $\zeta$(H) 
in a step of 0.5 dex. 
We specify the radiation field in terms of the standard 
Galactic Habing radiation field (G0). 
The shape of the SED is that given by \citet{1987ASSL..134..731B} but assuming
an extinction of 1 to 4 Ryd radiation as this is highly absorbed in ISM.
Based on the observed H${_3^+}$ and H$_2$ column density ($\pm$ 1$\sigma$) contours we estimate $\zeta$(H$_2$). 

In Section {\ref{subsec:effect}} we mention that the electron density 
depends on the presence of PAHs. 
PAHs have been detected in various places of our Galaxy with large abundances. 
However, there are some regions, like the ionized part of the Orion Bar 
\citep{1990ApJ...349..120S} where 
PAHs do not exist. In NGC 7023 NW \citep{2013A&A...552A..15M}, PAHs with less than 50 
Carbon atoms are not present.
Hence, we perform two sets of calculations for these sightlines. 
In one set we do not include PAHs while in the other set we include PAHs. 
\citet{2008ARA&A..46..289T} has determined the abundances of C locked 
up in PAHs containing 20--100 C atoms as 14 parts per million H atom. As the amount of PAHs is not exactly determined, 
we consider three different amount 
of PAHs depending on \citet{2008ARA&A..46..289T}, PAH$_{lo}$, PAH$_{avg}$, and PAH$_{hi}$. 
The number of PAHs per hydrogen for PAH$_{lo}$, PAH$_{avg}$, and PAH$_{hi}$ are 10$^{-6.85}$, 10$^{-6.52}$, 
and 10$^{-6.15}$, respectively.

\begin{deluxetable}{ccccccl}
\tablecaption{list of all the observed information for the nine  
sightlines modelled \label{tab:1}}
\tablehead{
\colhead{Objects} & \colhead{E(B-V)} & \colhead{E(B-V)} & \colhead{Density} &\colhead{Density} & \colhead{$N$(H$_3^+$) $\pm$ 1$\sigma$} & \colhead{$N$(H$_2$) $\pm$ 1$\sigma$} \\
\colhead{ } & \colhead{mag} & \colhead{Ref. } & \colhead{cm$^{-3}$} &\colhead{Ref.} & \colhead{10$^{13}$ cm$^{-2}$} & \colhead{10$^{20}$ cm$^{-2}$ }
}
\startdata
HD 169454 & 1.12  & 1  & 300 & 2 & 5.93 $\pm$ 0.34  & 16.60 $\pm$ 8.37\\
HD 110432 & 0.51  & 3 & 140 & 2 & 5.22 $\pm$ 0.17  & 4.37 $\pm$ 0.29 \\
HD 204827 & 1.11 & 1 & 450 & 2 & 19.00 $\pm$ 2.54 & 20.90 $\pm$ 10.20 \\
$\lambda$ Cep & 0.57 & 2 & 115 & 2 & 7.58 $\pm$ 1.17 & 6.88 $\pm$ 0.48 \\
X Per & 0.59 & 2  & 325 & 2 & 7.34 $\pm$ 0.92 & 8.38 $\pm$ 0.89\\
HD 73882 & 0.70 & 4 & 520  & 2 & 9.02 $\pm$ 0.50 & 12.90 $\pm$ 2.39 \\
HD 154368 & 0.78  & 4 & 240 & 2 & 9.37 $\pm$ 1.32  & 14.40 $\pm$ 3.99  \\
Cyg OB2 5 & 1.99 & 5 & 225 & 2 & 24.00 $\pm$ 3.29  & 15.20 $\pm$ 7.39 \\
Cyg OB2 12 & 3.35 & 5 & 300 & 2& 34.30 $\pm$ 5.89 & 80.00 $\pm$ 69.10 \\
\enddata

\tablecomments{Ref. (1) \citet{2003ApJ...584..339T}, (2) \citet{2007ApJS..168...58S}, 
(3) \citet{2002ApJ...577..221R}, (4) \citet{2009ApJS..180..125R}, (5) \citet{2002ApJ...567..391M}  }

\end{deluxetable}

\section{Results} \label{sec:results}
In this section, we present our results for nine sightlines using detailed 
numerical simulations.  We create contour plots showing column densities of H$_3^+$ and H$_2$ 
as functions of X$_{CR}$ and radiation field intensity (in terms of G0). 
Here, $\zeta(H)$ = 2$\times$X$_{CR}$ $\times$ 10$^{-16}$ s$^{-1}$. 

\begin{figure}[]
\includegraphics[scale=0.7]{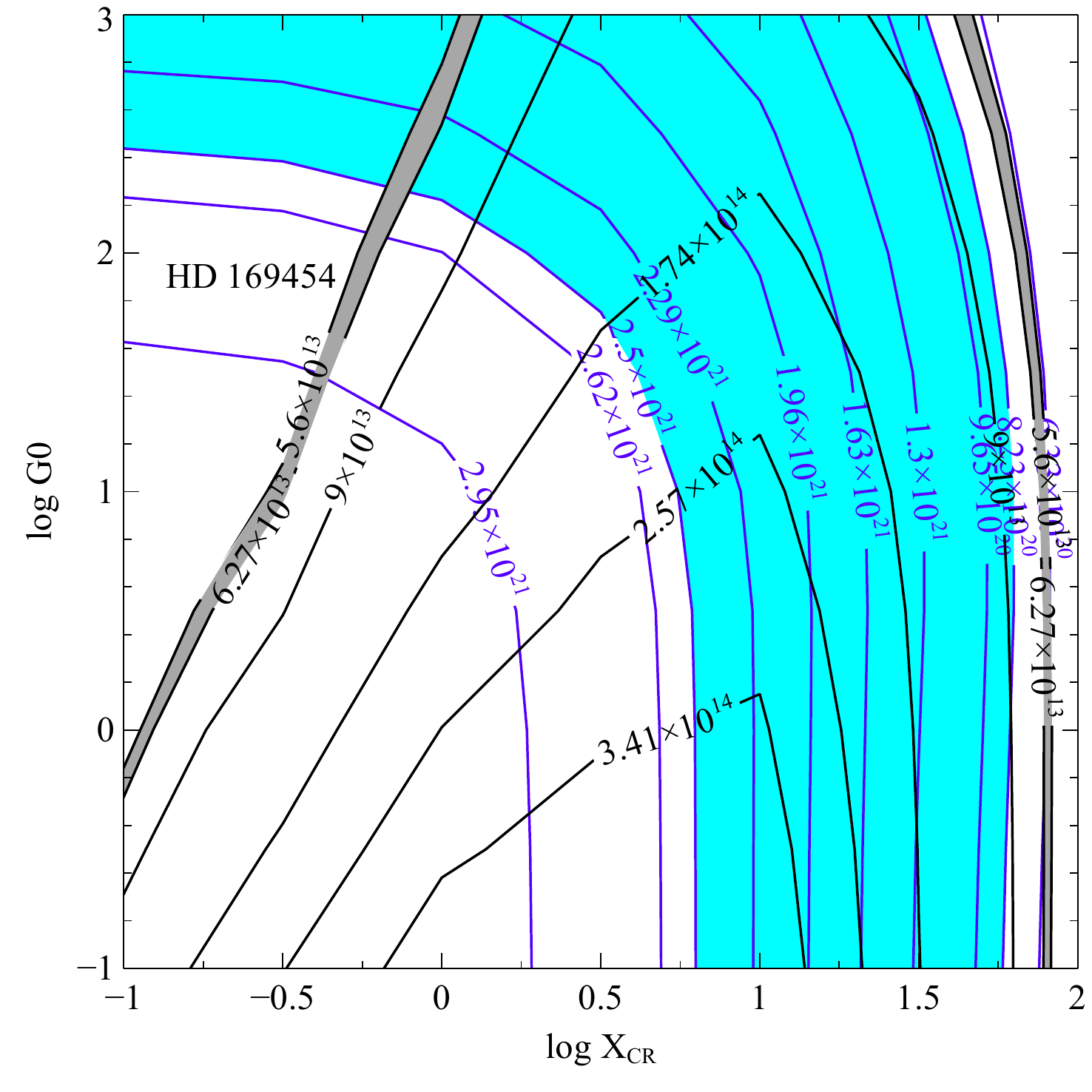}
\caption{Contour plot of H$_3^+$ and H$_2$ as a function of X$_{CR}$ and radiation field (in terms of G0) for HD 169454. The black and blue solid lines represent contour plots of column densities for H$_3^+$  and H$_2$, respectively. The filled areas represent observed column density values $\pm$ 1 $\sigma$.} \label{fig:model1}
\end{figure}

Fig. {\ref{fig:model1}}, Fig. {\ref{fig:model2}}, Fig. {\ref{fig:model3}}, Fig. {\ref{fig:model4}}, Fig. {\ref{fig:model5}}, Fig. {\ref{fig:model6}}, Fig. {\ref{fig:model7}}, Fig. {\ref{fig:model8}}, and Fig. {\ref{fig:model9}} show contour plots of H$_3^+$ and H$_2$ column densities as a function of X$_{CR}$ 
and radiation field (in terms of G0) for HD 169454, HD 110432, HD 204827, $\lambda$ Cep, X Per, HD 73882, HD 154368, Cyg OB2 5, 
Cyg OB2 12 without any PAHs, respectively. The black and blue solid lines represent contour plots of column densities for H$_3^+$  and H$_2$, respectively. 
Whereas, the filled areas with gray and cyan represent observed column density values, for H$_3^+$  and 
H$_2$, $\pm$ 1 $\sigma$. 

\begin{figure}[]
\includegraphics[scale=0.7]{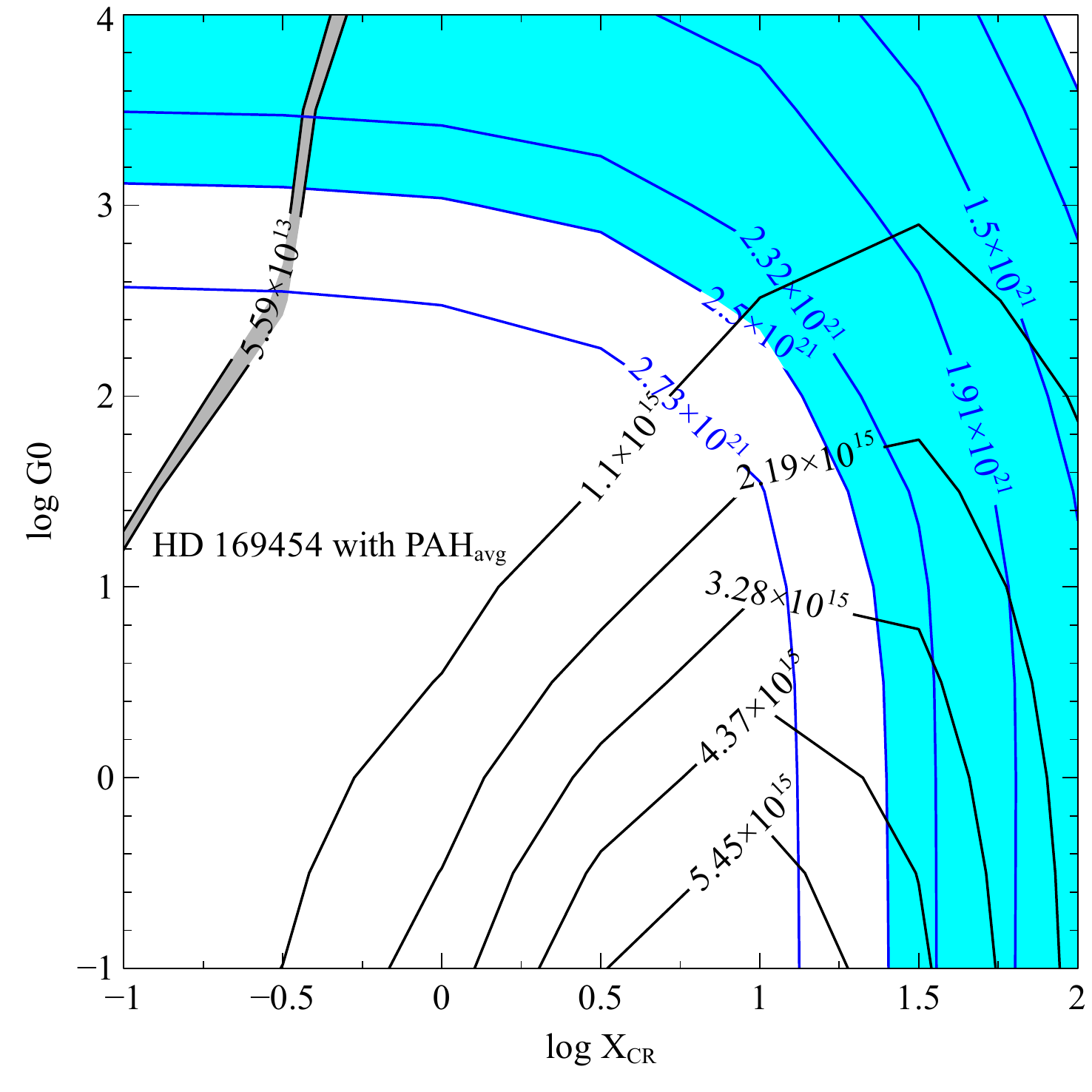}
\caption{Contour plot of H$_3^+$ and H$_2$ as a function of  X$_{CR}$ and radiation field (in terms of G0) for HD 169454 
with PAHs$_{avg}$. The black and blue solid lines represent contour plots of column densities for H$_3^+$  and H$_2$, respectively. The filled areas represent observed column density values $\pm$ 1 $\sigma$.}
\label{fig:model1pah9}
\end{figure}

\begin{figure}[]
\includegraphics[scale=0.7]{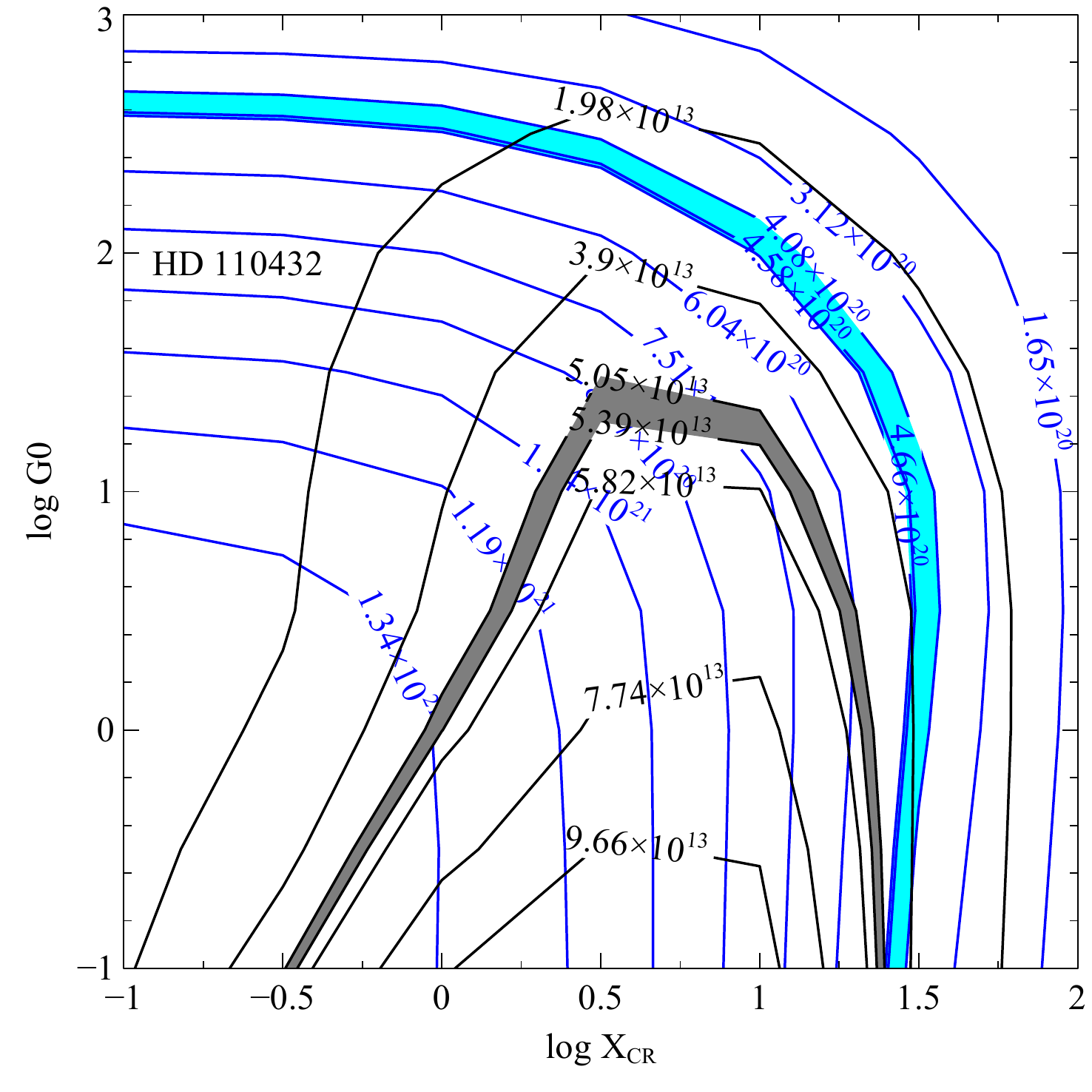}
\caption{ Contour plot of H$_3^+$ and H$_2$ as a function of X$_{CR}$ and radiation field (in terms of G0) for HD 110432. The black and blue solid lines represent contour plots of column densities for H$_3^+$  and H$_2$, respectively. The filled areas represent observed column density values $\pm$ 1 $\sigma$.}
\label{fig:model2}
\end{figure}

\begin{figure}[]

\includegraphics[scale=0.7]{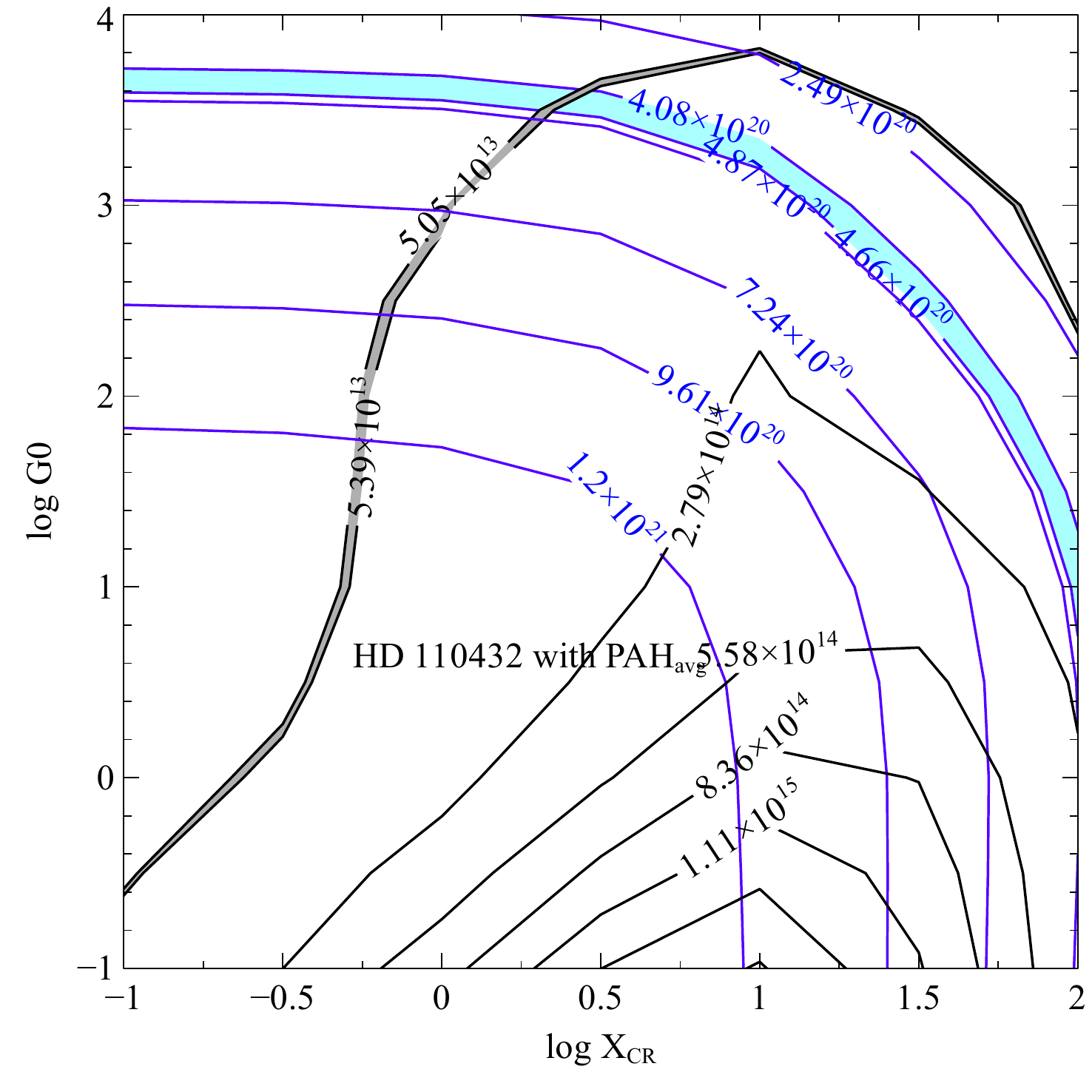}
\caption{ Contour plot of H$_3^+$ and H$_2$ as a function of  X$_{CR}$ and radiation field (in terms of G0) for HD 110432 with 
PAH$_{avg}$. The black and blue solid lines represent contour plots of column densities for H$_3^+$  and H$_2$, respectively. The filled areas represent observed column density values $\pm$ 1 $\sigma$.} 
\label{fig:model2pah9}
\end{figure}

\begin{figure}[]

\includegraphics[scale=0.7]{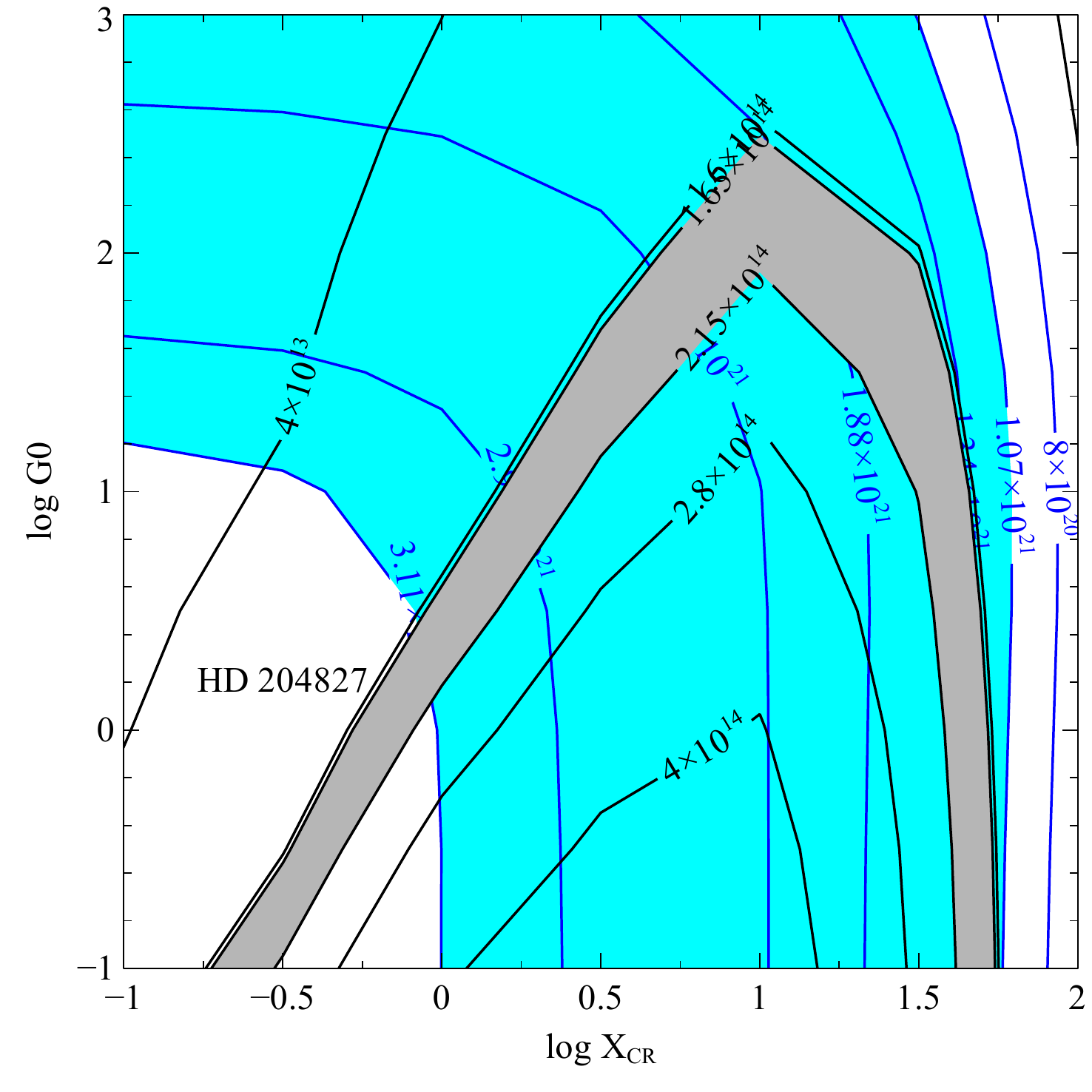}
\caption{Contour plot of H$_3^+$ and H$_2$ as a function of X$_{CR}$ and radiation field (in terms of G0) for HD 204827. The black and blue solid lines represent contour plots of column densities for H$_3^+$  and H$_2$, respectively. The filled areas represent observed column density values $\pm$ 1 $\sigma$.}
\label{fig:model3}
\end{figure}

\begin{figure}[]

\includegraphics[scale=0.7]{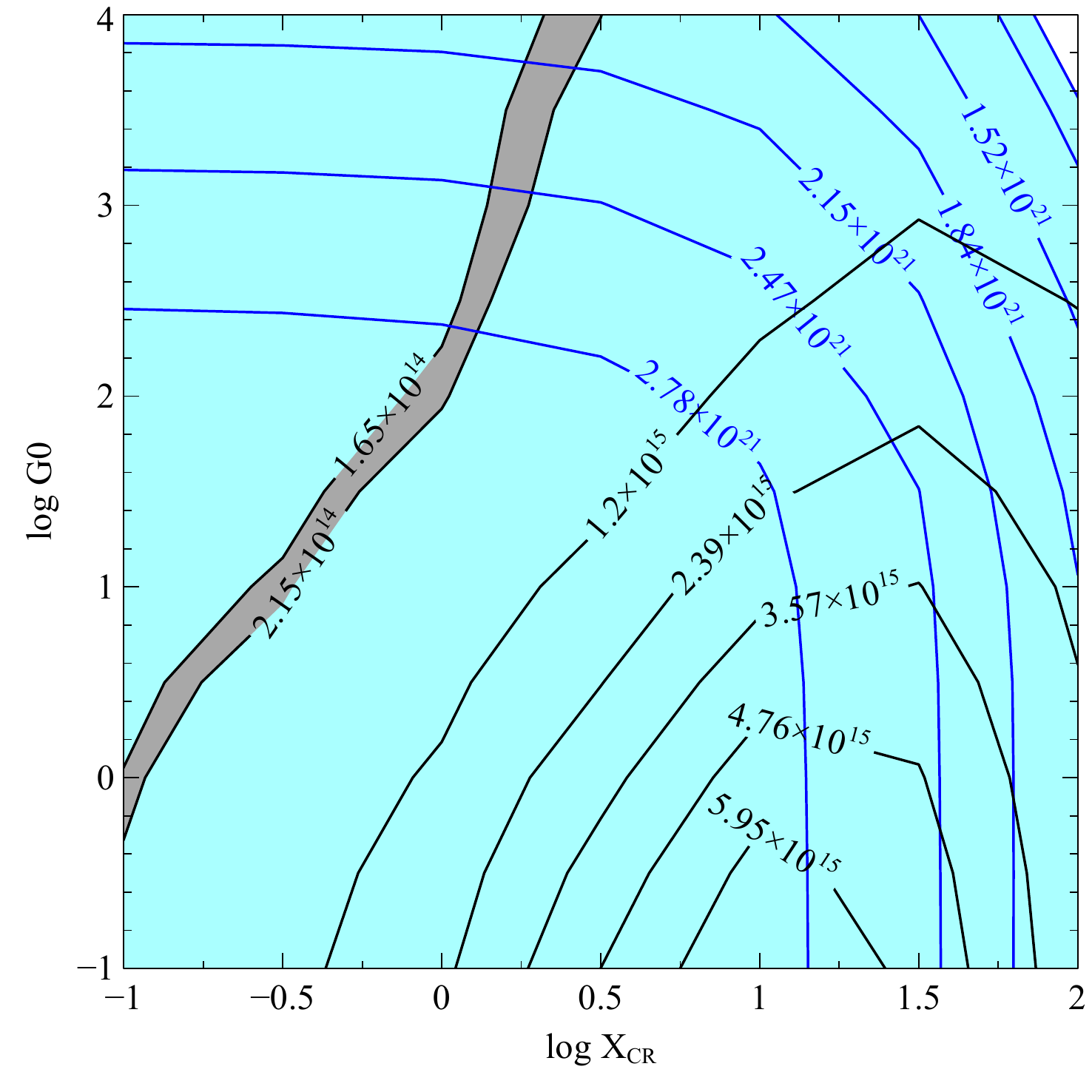}
\caption{Contour plot of H$_3^+$ and H$_2$ as a function of X$_{CR}$ and radiation field (in terms of G0) for HD 204827 
with PAH$_{avg}$. The black and blue solid lines represent contour plots of column densities for H$_3^+$  and H$_2$, respectively. The filled areas represent observed column density values $\pm$ 1 $\sigma$.}
\label{fig:model3pah9}
\end{figure}

\begin{figure}[]

\includegraphics[scale=0.7]{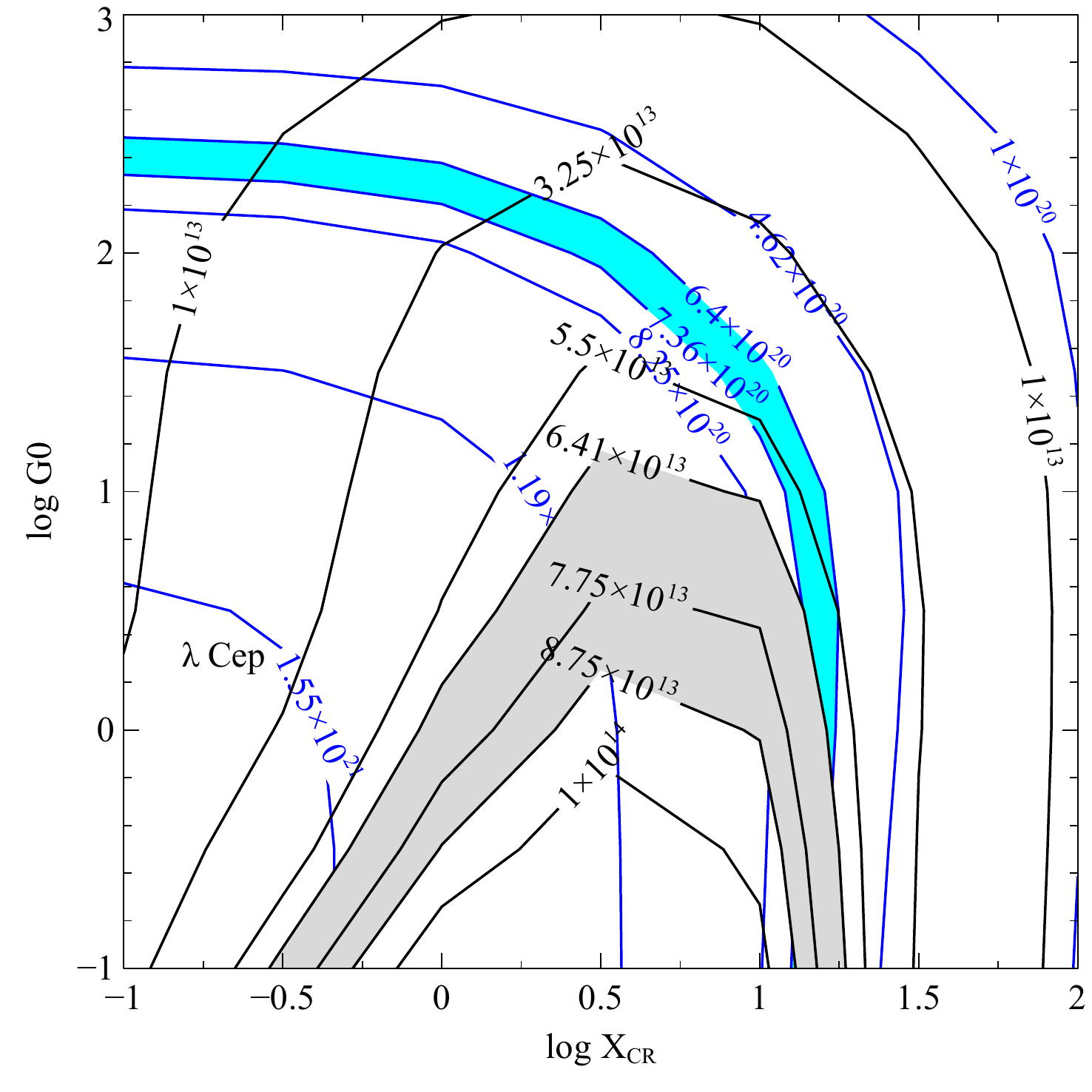}
\caption{Contour plot of H$_3^+$ and H$_2$ as a function of X$_{CR}$ and radiation field (in terms of G0) for $\lambda$ Cep. The black and blue solid lines represent contour plots of column densities for H$_3^+$  and H$_2$, respectively. The filled areas represent observed column density values $\pm$ 1 $\sigma$.}  
\label{fig:model4}
\end{figure}

\begin{figure}[]

\includegraphics[scale=0.7]{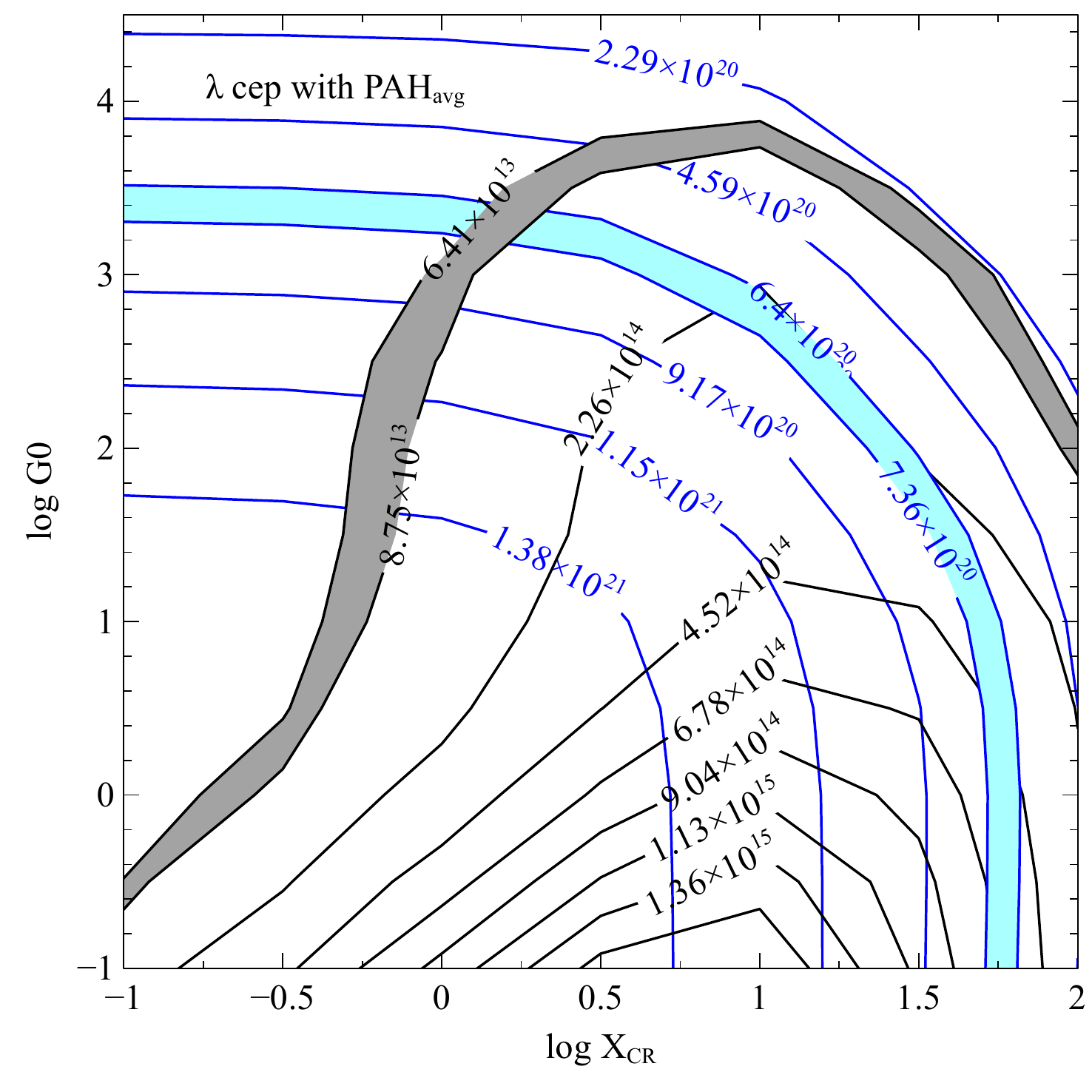}
\caption{Contour plot of H$_3^+$ and H$_2$ as a function of X$_{CR}$ and radiation field (in terms of G0) 
for $\lambda$ Cep with PAH$_{avg}$. The black and blue solid lines represent contour plots of column densities for H$_3^+$  and H$_2$, respectively. The filled areas represent observed column density values $\pm$ 1 $\sigma$.} \label{fig:model4pah9}
\end{figure}

\begin{figure}[]

\includegraphics[scale=0.7]{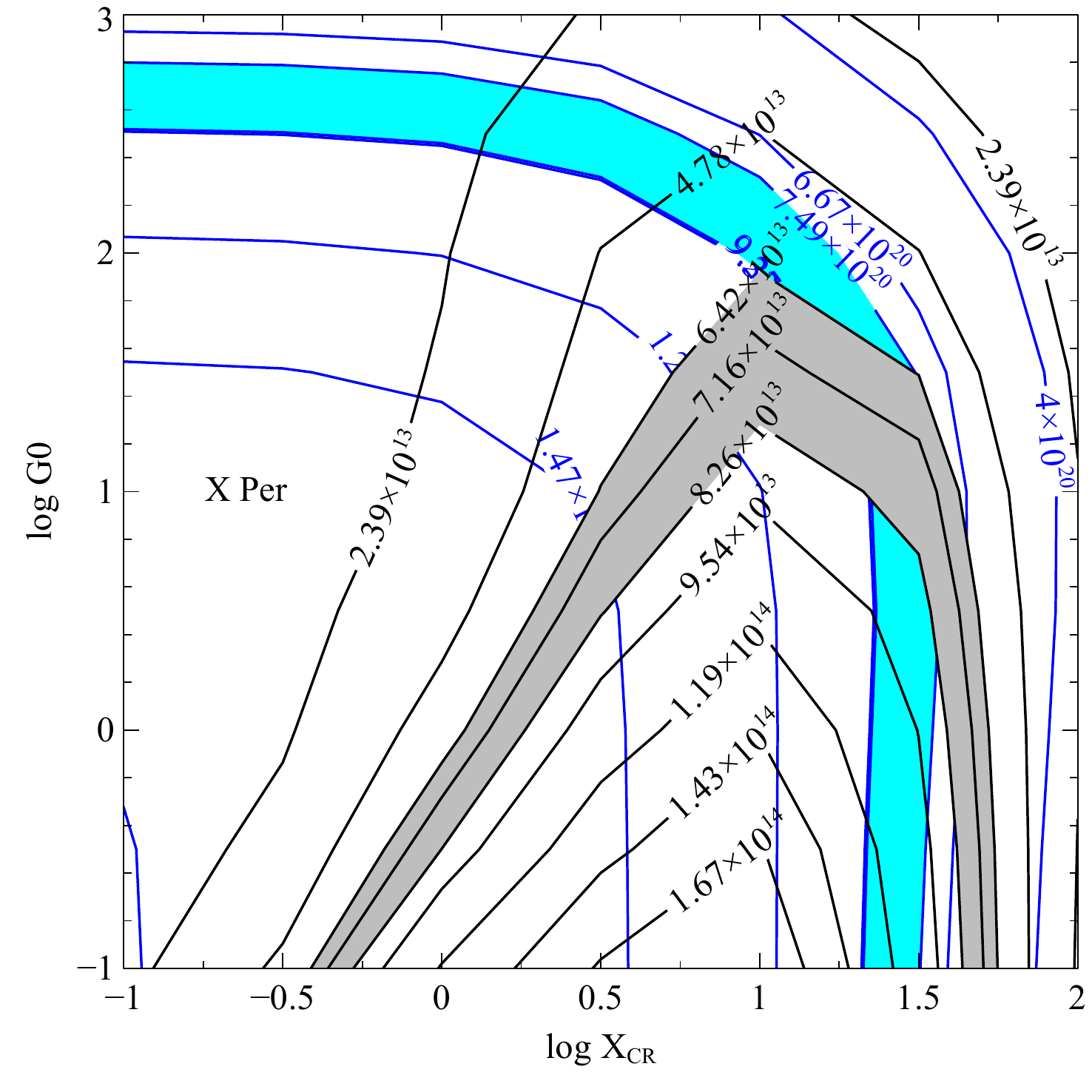}
\caption{Contour plot of H$_3^+$ and H$_2$ as a function of X$_{CR}$ and radiation field (in terms of G0) for 
X Per. The black and blue solid lines represent contour plots of column densities for H$_3^+$  and H$_2$, respectively. The filled areas represent observed column density values $\pm$ 1 $\sigma$.}
\label{fig:model5}
\end{figure}

\begin{figure}[]

\includegraphics[scale=0.7]{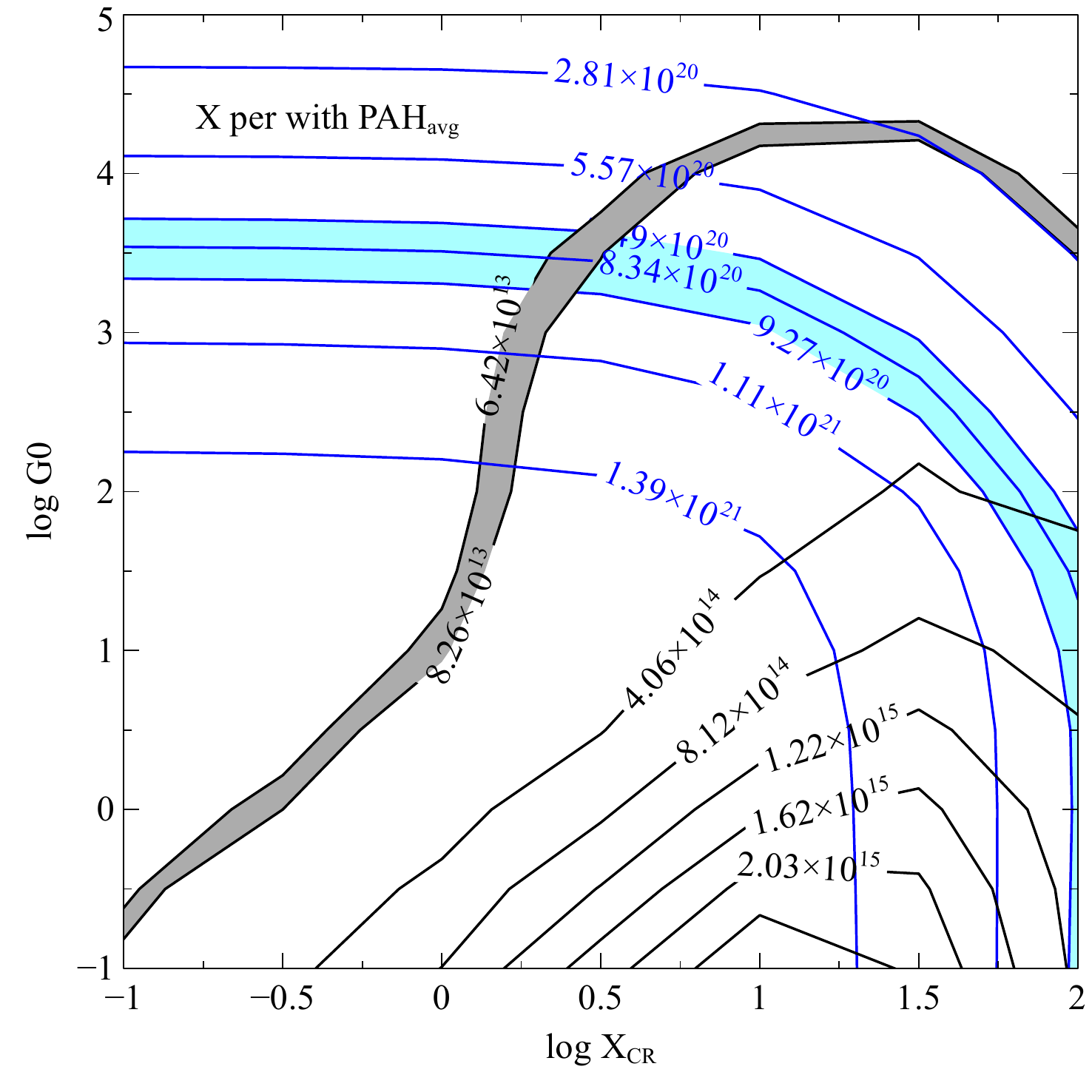}
\caption{Contour plot of H$_3^+$ and H$_2$ as a function of X$_{CR}$ and radiation field (in terms of G0) for X Per with PAH$_{avg}$. The black and blue solid lines represent contour plots of column densities for H$_3^+$  and H$_2$, respectively. The filled areas represent observed column density values $\pm$ 1 $\sigma$.}  
\label{fig:model5pah9}
\end{figure}

\begin{figure}[]

\includegraphics[scale=0.7]{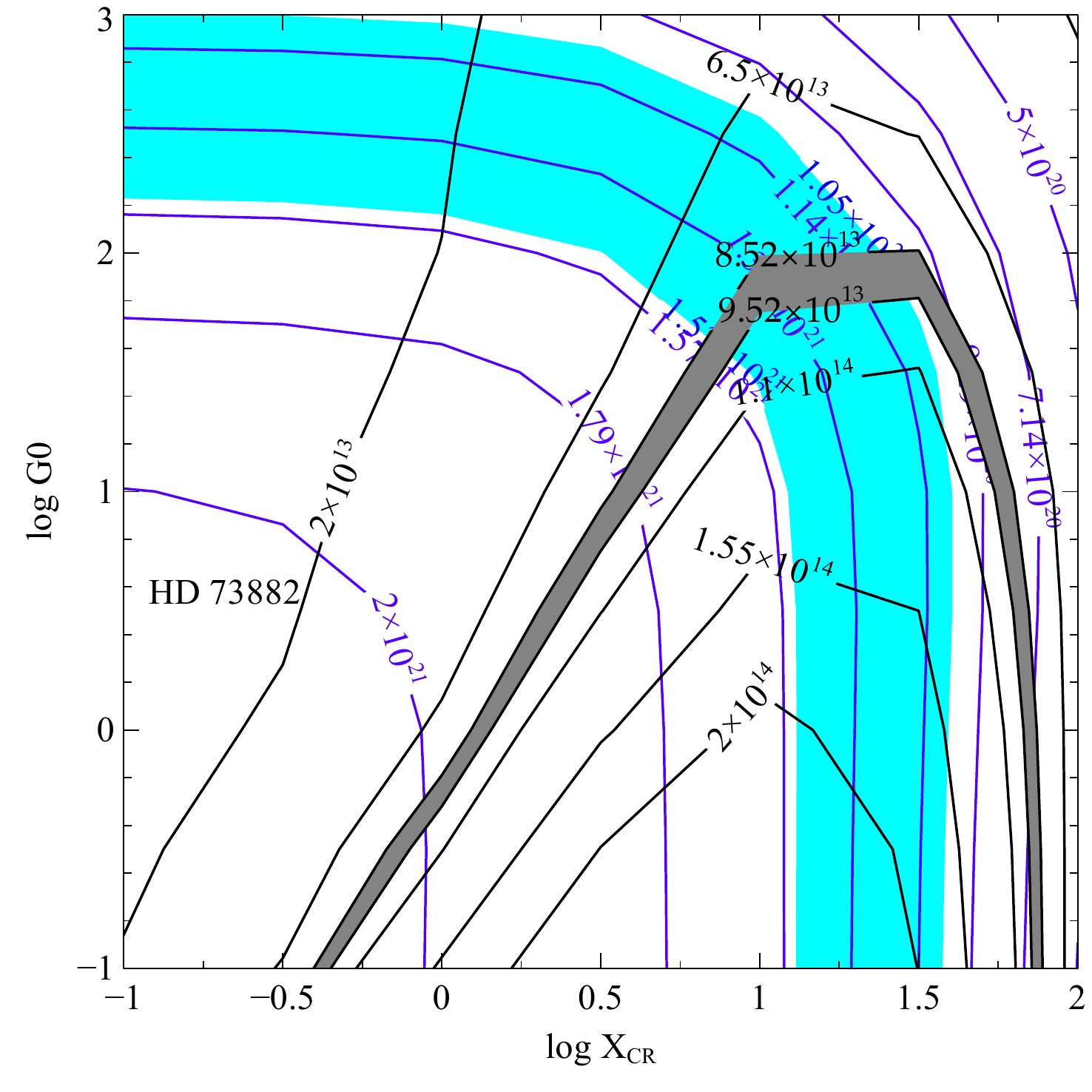}
\caption{Contour plot of H$_3^+$ and H$_2$ as a function of X$_{CR}$ and radiation field (in terms of G0) for HD 73882. The black and blue solid lines represent contour plots of column densities for H$_3^+$  and H$_2$, respectively. The filled areas represent observed column density values $\pm$ 1 $\sigma$.} 
\label{fig:model6}
\end{figure}

\begin{figure}[]

\includegraphics[scale=0.7]{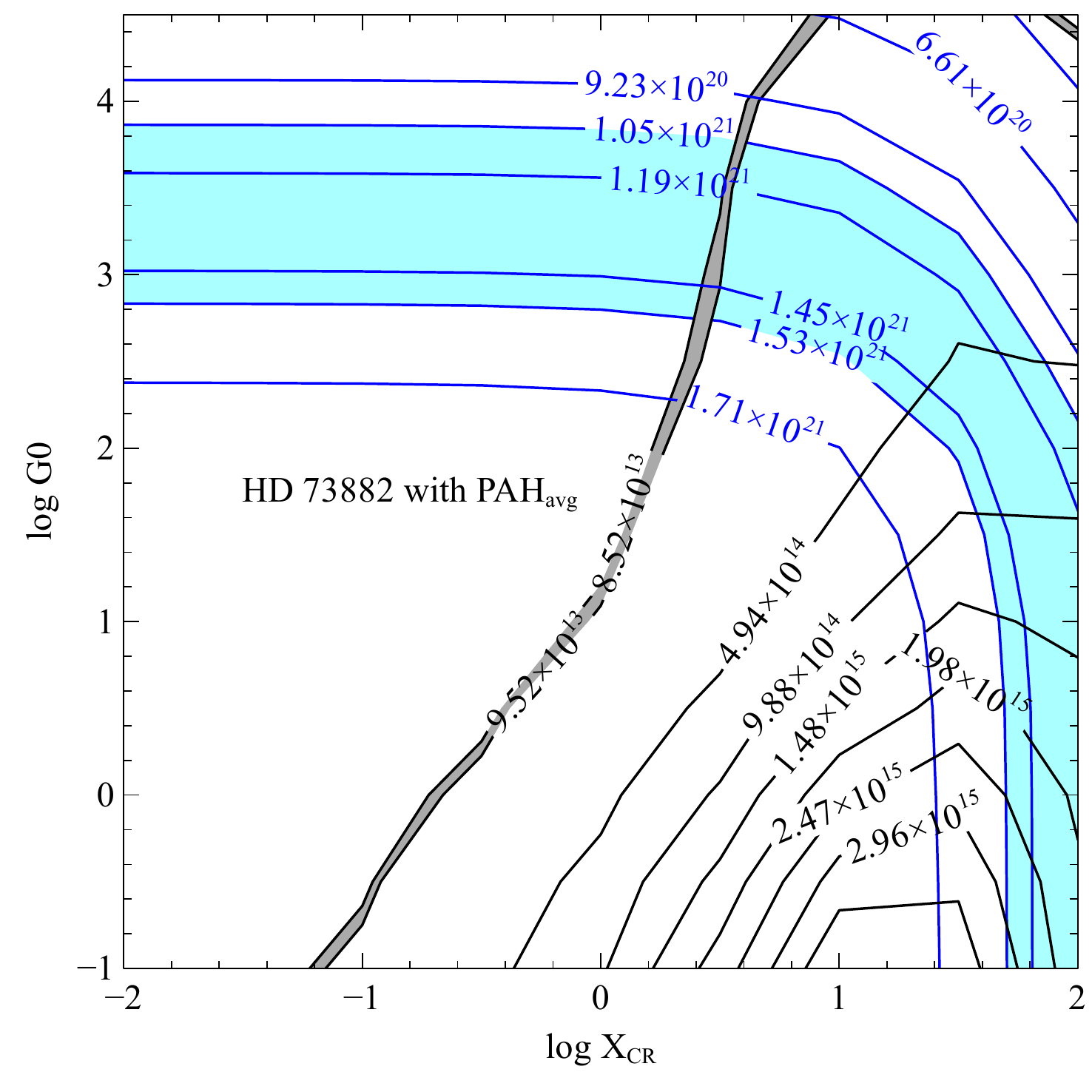}
\caption{Contour plot of H$_3^+$ and H$_2$ as a function of X$_{CR}$ and radiation field (in terms of G0) for HD 73882 
with PAH$_{avg}$. The black and blue solid lines represent contour plots of column densities for H$_3^+$  and H$_2$, respectively. The filled areas represent observed column density values $\pm$ 1 $\sigma$.} 
\label{fig:model6pah9}
\end{figure}

\begin{figure}[]

\includegraphics[scale=0.7]{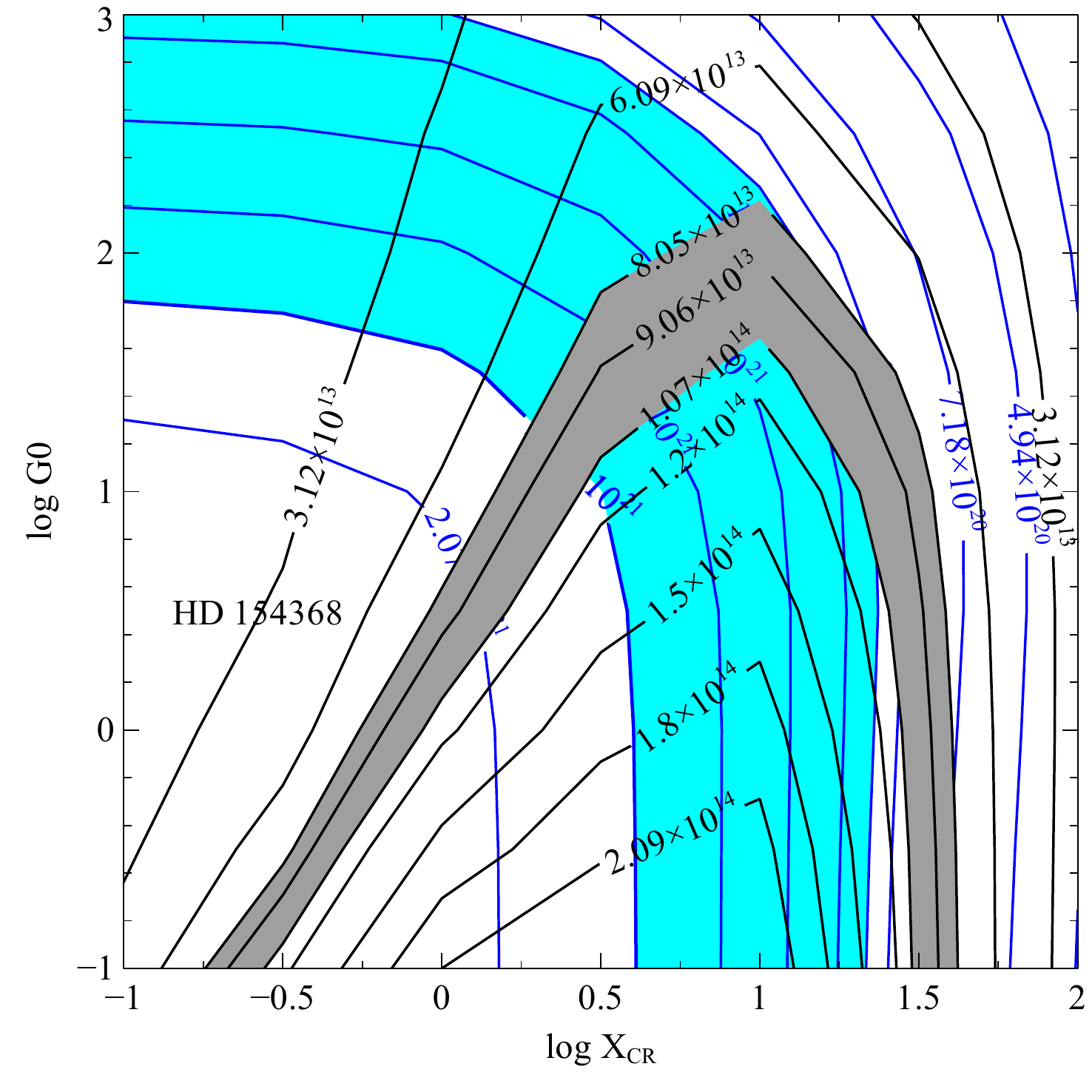}
\caption{Contour plot of H$_3^+$ and H$_2$ as a function of X$_{CR}$ and radiation field (in terms of G0) for HD 154368. The black and blue solid lines represent contour plots of column densities for H$_3^+$  and H$_2$, respectively. The filled areas represent observed column density values $\pm$ 1 $\sigma$.}
\label{fig:model7}
\end{figure}

\begin{figure}[]

\includegraphics[scale=0.7]{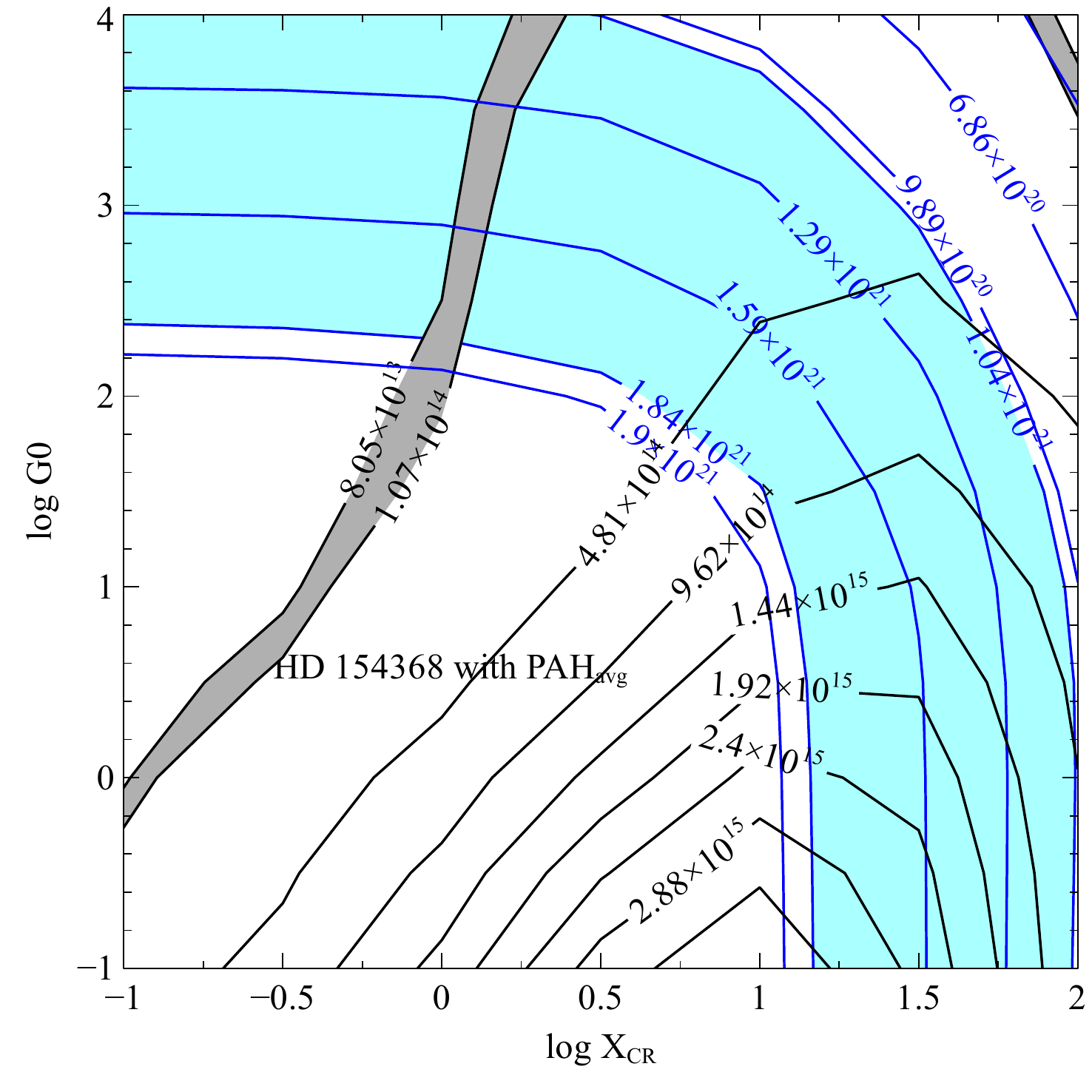}
\caption{Contour plot of H$_3^+$ and H$_2$ as a function of X$_{CR}$ and radiation field (in terms of G0) 
for HD 154368 with PAH$_{avg}$. The black and blue solid lines represent contour plots of column densities for H$_3^+$  and H$_2$, respectively. The filled areas represent observed column density values $\pm$ 1 $\sigma$.}
\label{fig:model7pah9}
\end{figure}

\begin{figure}[]

\includegraphics[scale=0.7]{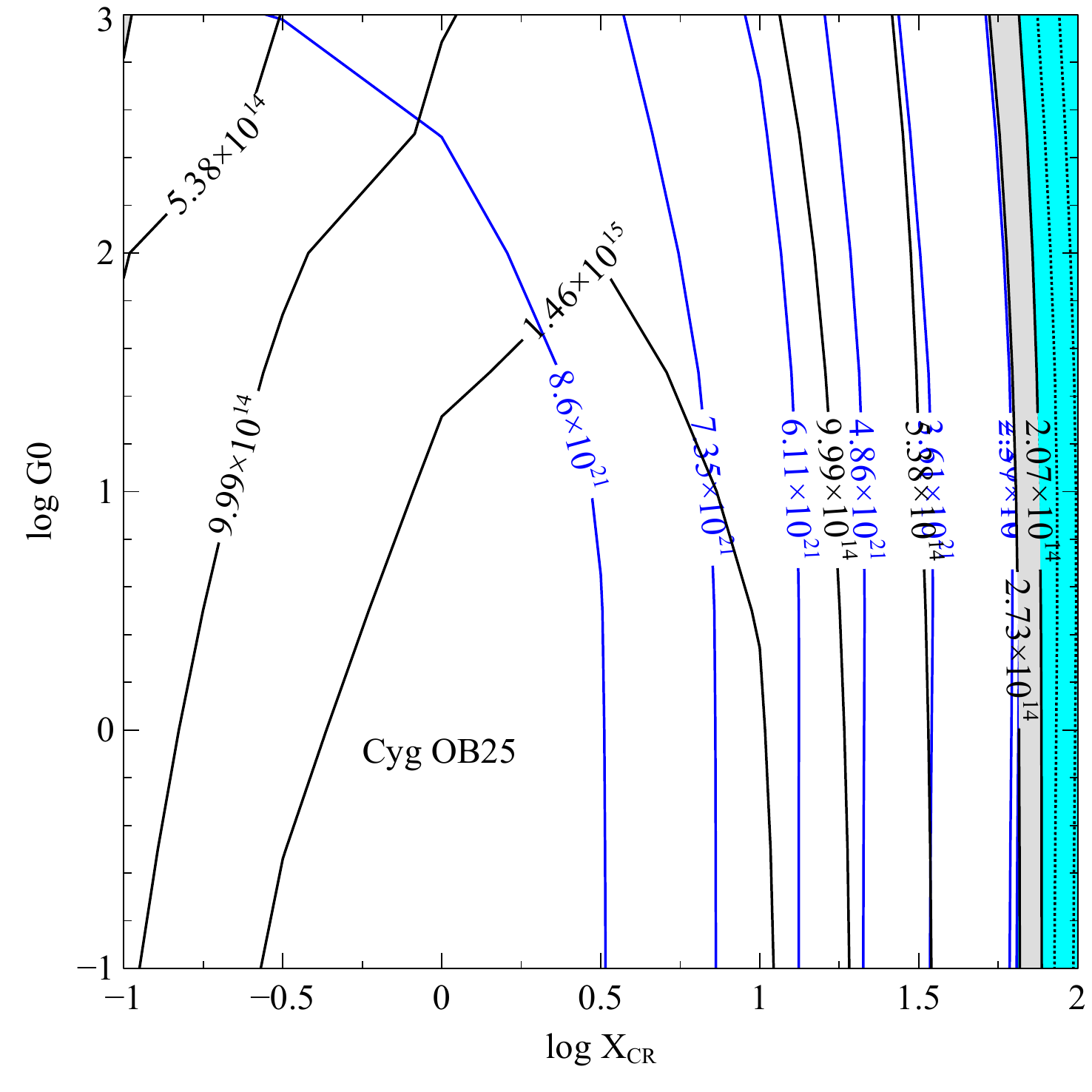}
\caption{Contour plot of H$_3^+$ and H$_2$ as a function of X$_{CR}$ and radiation field (in terms of G0) for Cyg OB2 5. The black and blue solid lines represent contour plots of column densities for H$_3^+$  and H$_2$, respectively. The filled areas represent observed column density values $\pm$ 1 $\sigma$.} 
\label{fig:model8}
\end{figure}

\begin{figure}[]

\includegraphics[scale=0.7]{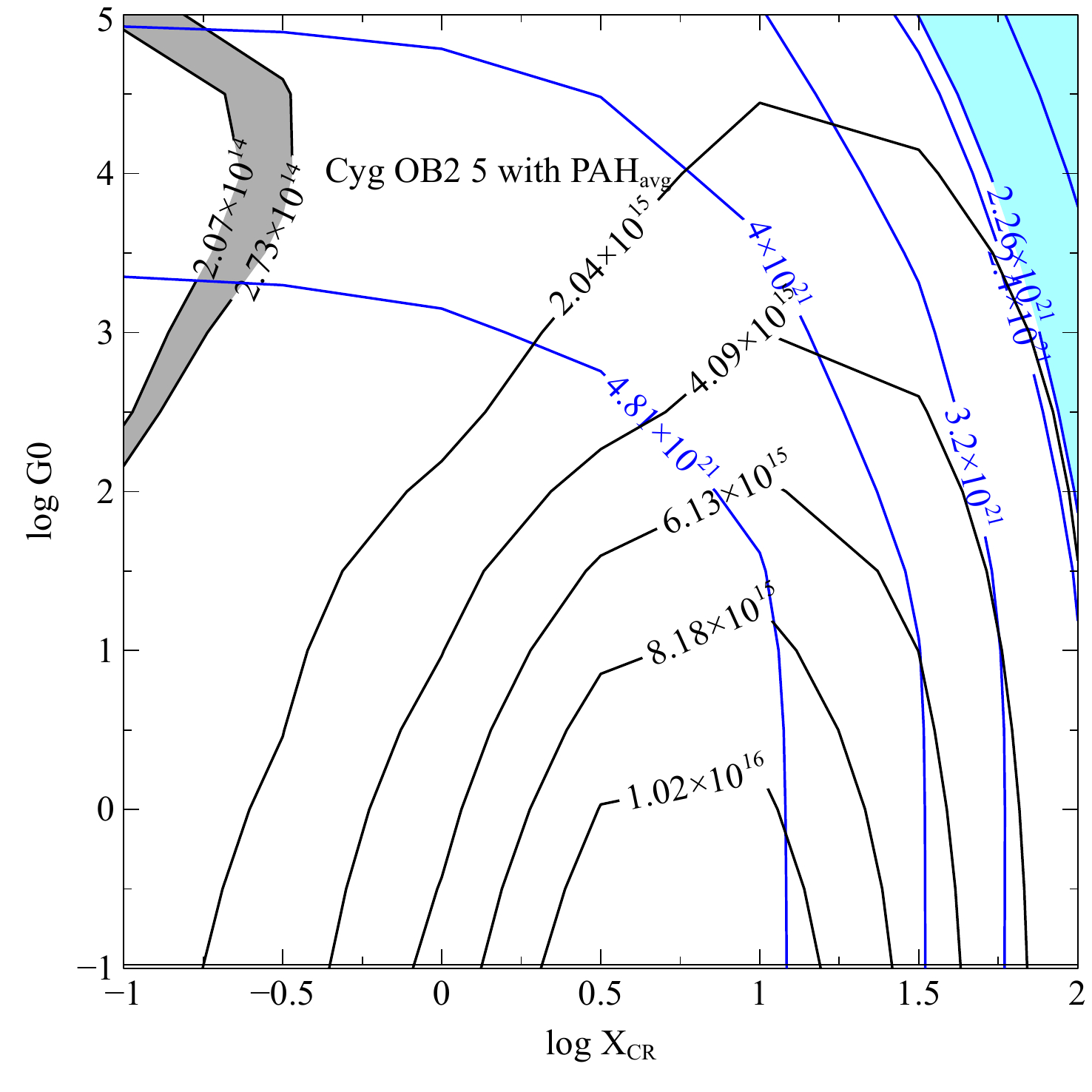}
\caption{Contour plot of H$_3^+$ and H$_2$ as a function of X$_{CR}$ and radiation field (in terms of G0) 
for Cyg OB2 5 with PAH$_{avg}$. The black and blue solid lines represent contour plots of column densities for H$_3^+$ and H$_2$, respectively. The filled areas represent observed column density  values $\pm$ 1 $\sigma$. 
In this case, they do not intersect in the considered parameter space.} 
\label{fig:model8pah9}
\end{figure}

\begin{figure}[]

\includegraphics[scale=0.7]{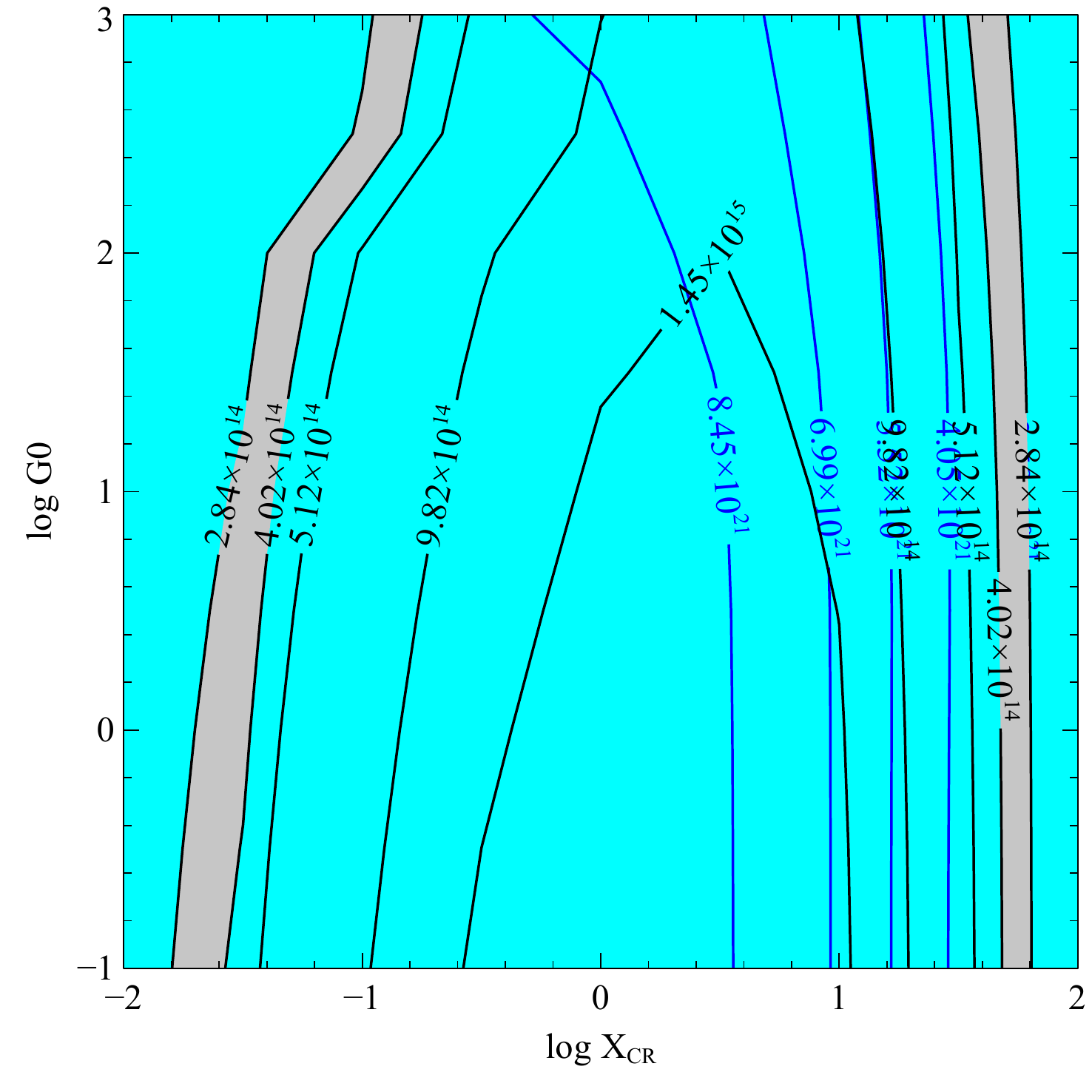}
\caption{Contour plot of H$_3^+$ and H$_2$ as a function of X$_{CR}$ and radiation field (in terms of G0) for Cyg OB2 12. The black and blue solid lines represent contour plots of column densities for H$_3^+$ and H$_2$, respectively. The filled areas represent observed column density values $\pm$ 1 $\sigma$.} 
\label{fig:model9}
\end{figure}

\begin{figure}[]

\includegraphics[scale=0.7]{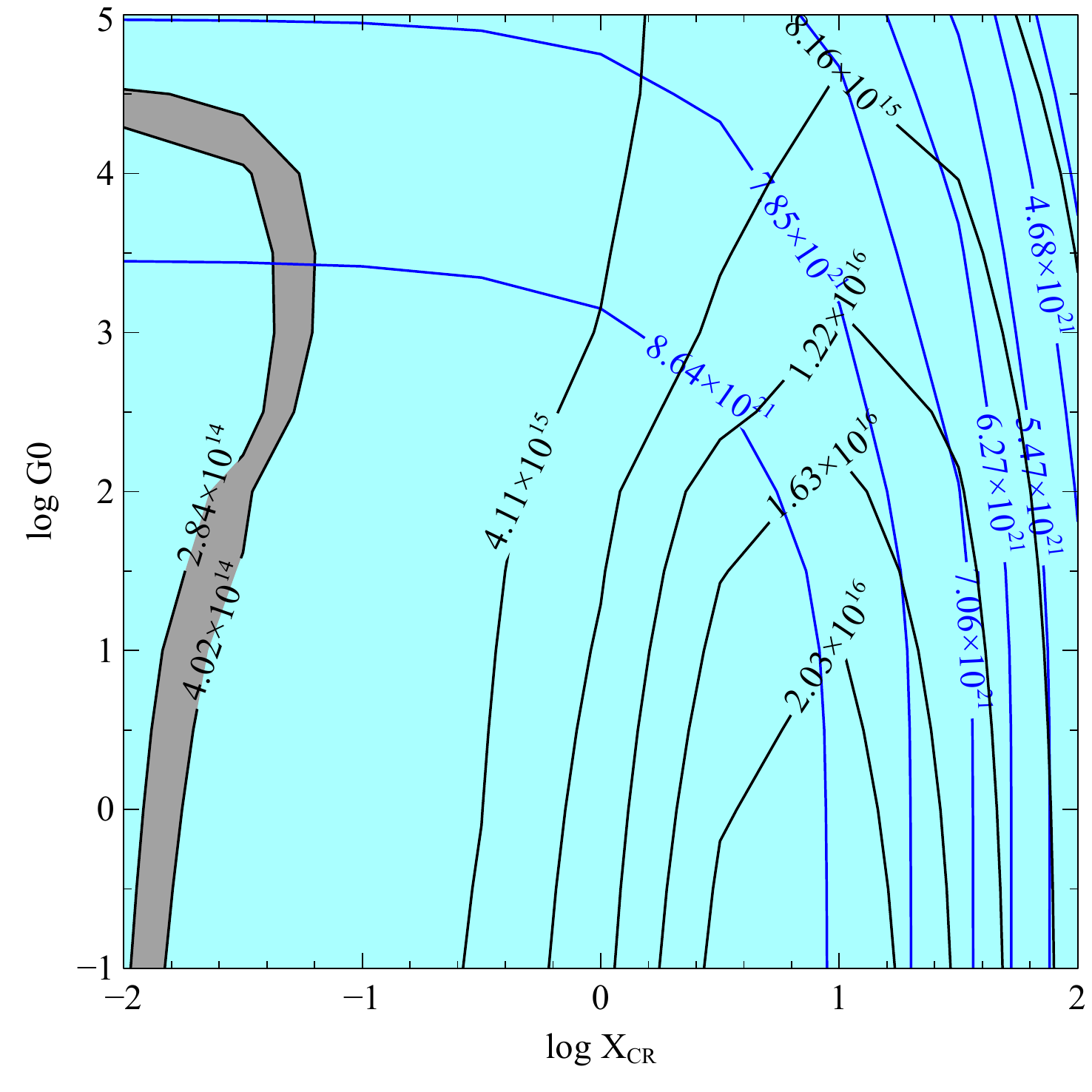}
\caption{Contour plot of H$_3^+$ and H$_2$ as a function of X$_{CR}$ and radiation field (in terms of G0) 
for Cyg OB2 12 with PAH$_{avg}$. The black and blue solid lines represent contour plots of column densities for H$_3^+$  and H$_2$, respectively. The filled areas represent observed column density values $\pm$ 1 $\sigma$.}  
\label{fig:model9pah9}
\end{figure}

In the same manner, Fig. {\ref{fig:model1pah9}}, Fig. {\ref{fig:model2pah9}}, Fig. {\ref{fig:model3pah9}}, Fig. {\ref{fig:model4pah9}}, Fig. {\ref{fig:model5pah9}}, Fig. {\ref{fig:model6pah9}}, Fig. {\ref{fig:model7pah9}}, Fig. {\ref{fig:model8pah9}}, and Fig. {\ref{fig:model9pah9}} show Contour plot of column densities for H$_3^+$ 
and H$_2$ as a function of X$_{CR}$ and radiation field (in terms of G0) with 
PAH$_{avg}$ for HD 169454, HD 110432, HD 204827, $\lambda$ Cep, X Per, HD 73882, HD 154368, Cyg OB2 5, Cyg OB2 12, respectively . Here also the black and blue solid lines represent contour plots of column densities for H$_3^+$  and H$_2$, respectively. 
Whereas, the filled areas with gray and cyan represent observed column density values, for H$_3^+$  and 
H$_2$, $\pm$ 1 $\sigma$.

Table {\ref{tab:2}} lists our findings. 
The second column of Table {\ref{tab:2}} shows the predicted value of $\zeta$(H$_2$) without considering PAHs in the
chemical network. It is clear that without PAHs, except for Cyg OB2 12, most of the sources have 
$\zeta$(H$_2$) $>$ 10$^{-16}$ s$^{-1}$. We estimate an average value of 
$\zeta$(H$_2$)= (95.69 $\pm$ 46.56) $\times$ 10$^{-16}$ s$^{-1}$ 
for models without PAHs.

The third column shows the predicted value of $\zeta$(H$_2$) with 
the lower PAH abundance, PAH$_{lo}$ in the chemical network. 
The presence of PAHs causes the electron density to decrease and the derived 
value of $\zeta$(H$_2$) to decrease. 
Adding PAHs causes the derived strength of the radiation field to increase.
We estimate an average value of 
$\zeta$(H$_2$)= (75.08 $\pm$ 39.90) $\times$ 10$^{-16}$ s$^{-1}$
for models with PAHs abundance, PAH$_{lo}$, except Cyg OB2 5 and Cyg OB2 12.

The fourth and fifth columns show the predicted value of $\zeta$(H$_2$) with 
the PAHs abundances PAH$_{avg}$ and PAH$_{hi}$, respectively.
As expected, the value of $\zeta$(H$_2$) decreases further and the derived radiation field also increases. 
Our contour plots not only provide an estimate of $\zeta$(H$_2$) but also an estimate of the prevailing radiation 
field in terms of G0. 
The estimated average values of 
$\zeta$(H$_2$) for these two cases are (7.88 $\pm$ 2.89) $\times$ 10$^{-16}$ s$^{-1}$ and 
(6.50 $\pm$ 3.06) $\times$ 10$^{-16}$ s$^{-1}$, except Cyg OB2 5 and Cyg OB2 12.
Column six lists the predicted values of $\zeta$(H$_2$) by \citet{2012ApJ...745...91I} for comparison.
Except for the two Cyg OB2 associations, the values of $\zeta$(H$_2$) matches with that 
predicted by \citet{2012ApJ...745...91I} within the error bars when PAH$_{avg}$ are considered. We predict 
$\approx$ 1 dex smaller $\zeta$(H$_2$) for the Cyg OB2 12 association than predicted by  \citet{2012ApJ...745...91I}.

\begin{deluxetable}{cccccl}
\tablecaption{Cosmic ray ionization rate \label{tab:2}}
\tablehead{
\colhead{} & \colhead{$\zeta$(H$_2$)} & \colhead{$\zeta$(H$_2$)} & \colhead{$\zeta$(H$_2$)} & \colhead{$\zeta$(H$_2$)} & \colhead{$\zeta$(H$_2$) $\pm$ 1$\sigma$}\\
\colhead{Objects } & \colhead{this work} & \colhead{this work}& \colhead{this work} & \colhead{this work} & \colhead{\citet{2012ApJ...745...91I}}\\
\colhead{ } & \colhead{no PAHs} &  \colhead{PAH$_{lo}$}& \colhead{PAH$_{avg}$} & \colhead{PAH$_{hi}$} & \colhead{ }\\
\colhead{ } & \colhead{10$^{-16}$ s$^{-1}$} & \colhead{10$^{-16}$ s$^{-1}$} & \colhead{10$^{-16}$ s$^{-1}$} & \colhead{10$^{-16}$ s$^{-1}$} & \colhead{10$^{-16}$ s$^{-1}$}
}

\startdata
HD 169454 & 2.64 to 5.15 & 2.41 to 4.47  &1.33 to 2.00 & 0.65 to 0.87 & 2.45 $\pm$ 1.83 \\
HD 110432 & $\approx$ 100.48 &120.80 to 126.49 & 7.98 to 11.80 & 10.04 to 14.86 & 3.86 $\pm$ 2.10 \\
HD 204827 & 3.73 to 224.93 & 2.85 to 276.73 & 0.40 to 12.65 &0.15 to 4.92 & 9.32 $\pm$ 6.92\\
$\lambda$ Cep & 50.35 to 63.40 &59.16 to 76.22 & 4.84 to 8.36 &4.81 to 8.55 & 2.84 $\pm$ 1.61 \\
X Per & 30.08 to 123.61 &29.65 to 152.08  & 7.57 to 14.86 &6.49 to 13.55 & 5.85 $\pm$ 3.54 \\
HD 73882 & 28.06 to 111.44 & 24.10 to 89.55 & 9.59 to 15.92 & 0.17 to 13.93 & 9.71 $\pm$ 5.57\\
HD 154368 & 7.80 to 89.55 & 6.79 to 79.81 & 3.41 to 9.59 & 1.75 to 10.28  & 4.19 $\pm$ 2.62 \\
Cyg OB2 5 & 219.00 to 303.43 &39.08 to 209.92 & -- & -- & 8.13 $\pm$ 6.03 \\
Cyg OB2 12 & 0.06 to 258.26  & 0.04 to 317.73 & 0.04 to 0.25 &0.04 to 0.22  & 2.93 $\pm$ 3.04  \\
\enddata

\end{deluxetable}

\begin{figure}[]

\includegraphics[scale=0.7]{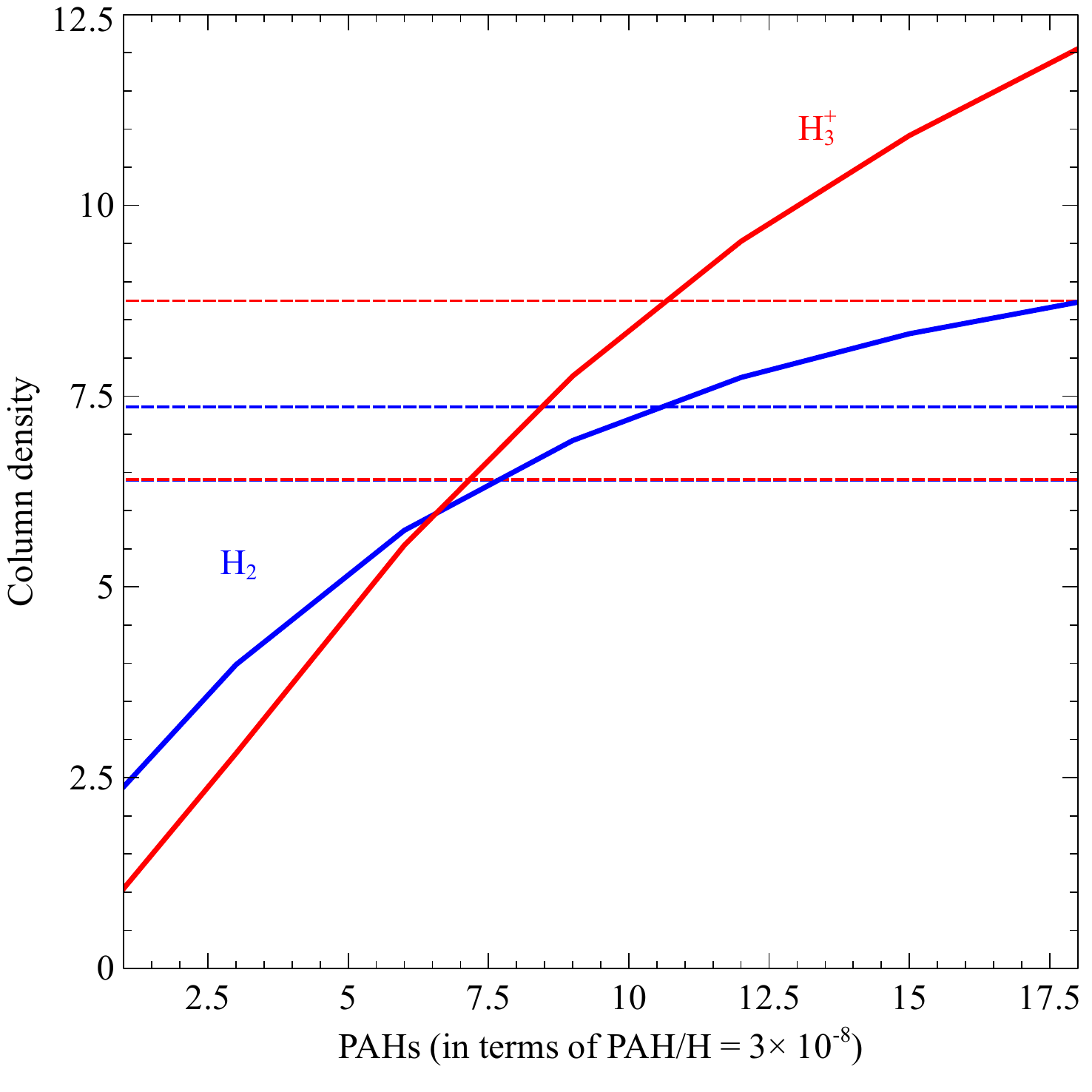}
\caption{Effect of PAHs on the column densities of  H$_3^+$ and H$_2$ for the sightline 
towards $\lambda$ Cep. The red and blue solid lines represent the predicted column densities for H$_3^+$ and H$_2$, respectively. The red and blue dashed-lines represent respective observed column density values $\pm$ 1 $\sigma$. 
Here H$_3^+$ and H$_2$ column densities are scaled by 10$^{13}$ and 10$^{20}$, respectively.} 
\label{fig:model4vary}
\end{figure}

\section {Discussions} \label{sec:disc}
\citet{2017ApJ...845..163N} have also performed detail numerical simulations of 7 sightlines based 
on observed column densities of H$_3^+$  and H$_2$ and reported an average value 
of (5.3$\pm$1.1) $\times$ 10$^{-16}$ s$^{-1}$. 
They have included PAHs in their calculations but 
didn't vary their abundance. 
It is very important to estimate the amount of PAHs since the value of the observed 
column densities of H$_3^+$ and H$_2$ and $\zeta$ depends on it significantly. 
To illustrate this, we examine the effects of a range of PAH abundances 
on the molecular abundances in the cloud along the sightline to
$\lambda$ Cep.
As a baseline model we choose 
the radiation field (10$^{3.3}$ G0) and $\zeta$(H$_2$) (6.0 $\times$ 10$^{-16}$ s$^{-1}$),
one of the possible solutions for the PAH$_{avg}$ case described above.
Fig.{\ref{fig:model4vary}} shows the results of varying the PAH abundance.
The red and blue solid lines represent the predicted column densities for H$_3^+$ and H$_2$, respectively. 
The red and blue solid dashed-lines represent respective observed column density values $\pm$ 1 $\sigma$. 
It is clear that a factor of 2 change in the PAHs abundances 
will predict column densities beyond observed range.

We notice that HD 169454, HD 204827, and the Cyg OB2 12 associations which 
have higher E(B-V) than the rest of the sample, predicts smaller value of $\zeta$(H$_2$) than the 
rest with high PAHs abundances, PAH$_{hi}$.
A similar dependence on A$_V$ was also reported by \citet{2017ApJ...845..163N}. 
Though we can compare the predicted values of $\zeta$(H$_2$), observed values of the corresponding radiation 
field are not available. Our typical derived values of G0 significantly higher than
 the ISM background, typically G0 10 to 500, consistent with the 
 clouds being near bright stars.

\section {Summary} \label{sec:summary}

We have used the spectral synthesis code CLOUDY \citep{2017RMxAA..53..385F} 
to create detailed simulations of the nine sightlines towards 
HD 169454, HD 110432, HD 204827, $\lambda$ Cep, X Per, HD 73882, HD 154368, Cyg OB2 5, 
Cyg OB2 12 and  determine the value of $\zeta$(H$_2$) based on the observed H$_3^+$  and H$_2$ 
column densities $\pm$ 1 $\sigma$. 
Our goal is to determine the cosmic-ray ionization rate.
We also check how sensitive the derived value is to physical assumptions
since, as pointed out by \citet{2006PNAS..10312269D}, there are many details in the chemistry
that affect the derived quantities.

The electron density plays a pivotal role in determining $\zeta$(H$_2$). 
The physical structure of the clouds along these sightlines is similar to a PDR. 
Hence we first study the electron density and its 
dependence on various parameters for a standard PDR model, model F1
of the 2006 Leiden PDR workshop \citet{2007A&A...467..187R}. 
Then we perform detailed 
grid calculations for these nine sightlines.
The cloud densities are derived from observed C$_2$ levels \citep{{2007ApJS..168...58S},{2002ApJ...577..221R}} 
and the cloud thicknesses are set from the observed E(B-V). 

Very small grains or large molecules can become charged and
affect the free electron density and derived cosmic-ray ionization rate.
Although PAHs are very important, their presences or absence is difficult to 
determine from absorption-line data and we know that there are some regions, 
the ionized part of the Orion Bar 
and NGC 7023 NW, where PAHs and PAHs with less than 50 C atoms are not present, respectively. 
Hence, to study the effect of PAHs on the derived cosmic-ray ionization rate, we consider 
four separate cases, 
without PAHs and with three values of the PAH abundances. 
We then solve for the cosmic-ray ionization rate and radiation field intensity. 
The values of $\zeta$ for these sightlines differ significantly with and without PAHs.

Our main conclusions from this work are listed below:
\begin{itemize}

\item The common assumption that all  carbon is  in C$^+$  in regions
where H$_3^+$ forms is not true in detail. 
Carbon is mostly  C$^{+}$ at shallower A$_V$ 
but becomes neutral or molecules as A$_V$ increases (See Fig.{\ref{fig:fig1}}). 
\item The common  assumption that all  electrons are contributed by 
C$^+$ where H$_3^+$ forms is not true in detail. 
The total electron density is affected by ionization of H$^+$, S$^+$, 
and other metals, and by the effects of PAHs.
Hence, e$^-$/H is not equal to C/H. 
For instance, we find that, for the Leiden F1 model, 
$n($e$^-$)/$n($H) $>$ $n($C)/$n($H) (See Fig. {\ref{fig:fig1}}).
\item We show that the electron density depends on the radiation field, 
the  presence of big molecules (PAHs) or very small grains, and $\zeta$.
This affects the H$_3^+$ abundance and derived $\zeta$(H$_2$).

\item  We predict an average cosmic-ray ionization rate 
 $\zeta$(H$_2$) equal to (75.08 $\pm$ 39.90) $\times$ 10$^{-16}$ s$^{-1}$, (7.88 $\pm$ 2.89) $\times$ 10$^{-16}$ s$^{-1}$, 
 and (6.50 $\pm$ 3.06) $\times$ 10$^{-16}$ s$^{-1}$ for 
 our PAH$_{lo}$, PAH$_{avg}$, and PAH$_{hi}$ cases, respectively, except the two sightlines 
towards the Cyg OB2 associations.
\item We estimate an average value of $\zeta$(H$_2$)= (95.69 $\pm$ 46.56) $\times$ 10$^{-16}$ s$^{-1}$ 
for models without PAHs. 
 \item Our derived value of $\zeta$ is nearly 1 dex smaller for 
the sightline towards Cyg OB2 12 than the value 
predicted by \citet{2012ApJ...745...91I} with PAH$_{avg}$ 
and PAH$_{hi}$.
A much higher rate, approaching the mean of the previous sightlines, is derived
when PAHs are not included.
This sightline has a highly uncertain H$_{2}$ column density. The value of $\zeta$(H$_{2}$) has
 an uncertainty of more than a dex due to this uncertainty. 

 
 
\end{itemize}
\acknowledgments

GS acknowledges WOS-A grant from Department of Science and Technology (SR/WOS-A/PM-9/2017). 
GJF acknowledges support by NSF (1816537, 1910687), NASA (ATP 17-ATP17-0141, 19-ATP19-0188), and STScI (HST-AR- 15018).
We thank the anonymous referee for his/her thoughtful suggestions.

\bibliography{COSMIC}{}
\bibliographystyle{aasjournal}
\end{document}